\documentclass[american,superscriptaddress,aps,prd,preprint,floatfix]{revtex4}
\usepackage[latin1]{inputenc}
\usepackage{float}
\usepackage{amsmath}
\usepackage{graphicx}
\usepackage{amssymb}
\usepackage{cancel}
\usepackage{epsfig}
\usepackage[dvips]{color}
\usepackage{subfigure}
\usepackage{psfrag}

\usepackage[bookmarks=true]{hyperref}
\usepackage{slashed}
\usepackage{lscape}

\usepackage{babel}
\makeatother
\begin{document}

\title{Step-scaling with off-shell renormalisation}

\newcommand\edinb{SUPA, School of Physics, The University of Edinburgh,
                  Edinburgh EH9 3JZ, UK}

\author{R. Arthur}     \affiliation{\edinb}
\author{P. A. Boyle}     \affiliation{\edinb}

\collaboration{RBC and UKQCD Collaborations}
\noaffiliation{Edinburgh-2010/14}

\date{\today}

\begin{abstract}
We make use of twisted boundary conditions
for off-shell, Rome-Southampton renormalisation.
This allows us to define vertex functions at a fixed physical momentum that need not be a Fourier
mode. This definition includes choosing a fixed
orientation with respect to the lattice axes; only then do
lattice artefacts, which include $O(4)$ breaking effects,
have a valid expansion in powers of the lattice spacing, $a$. 
The use of non-exceptional momenta has been found to greatly reduce
the dependence of the vertex functions on both mass and momentum, $p$.
Excellent statistical precision is afforded
by plane-wave sources. Together this enables both a theoretically well founded 
and statistically clean continuum limit. Thereafter all $p^2$ dependence
can be identified with the anomalous running of the operator.
We present initial results and develop a practical scheme
for step-scaling with off-shell renormalization.
The size of the steps is continuous rather than discrete, allowing
arbitrarily small steps and the scheme is easy to implement for general operators .
\end{abstract}

\maketitle

\section{Introduction}

The non-perturbative renormalisation (NPR) of Lattice QCD matrix elements
is an important topic for the phenomenological relevance 
of the field. This paper addresses two key limitations
of the popular Rome-Southampton RI/MOM method\cite{Martinelli:1994ty}: firstly 
the entangled discretisation and perturbative truncation errors and secondly the rather low scale 
at which continuum perturbation theory is applied.

Phenomenology requires that
scheme dependent QCD observables calculated within lattice QCD
are converted to a perturbatively amenable scheme such as $\overline{MS}$. 
This is often done using an intermediate, regularisation invariant
momentum scheme (known as RI/MOM). The conversion of lattice
results to the RI/MOM scheme is non-perturbative and 
the scale is defined by the momenta used. Obtaining
precision in the lattice calculation
requires that the momentum scale used is well below the lattice cut-off, 
giving the high end of the Rome-Southampton momentum ``window'' condition:
\begin{equation}
\label{eq:window}
\Lambda_{\rm QCD}^2 \ll p^2 \ll \left(\frac{\pi}{a}\right)^2.
\end{equation}
The perturbative conversion between 
RI/MOM (or RI/SMOM in the case of non-exceptional
kinematics) and $\overline{MS}$ is known for many operators
to as many as three loops. Perturbative accuracy is dependent on satisfying the
lower inequality in Eq. \eqref{eq:window}.
The error associated with the low end of the window 
depends on logarithms the physical momentum scale, while the
high end converges more benignly as a power of the dimensionless momentum scale
($(a p)^2$ in the case of domain wall fermions).
The continuum extrapolation enabled by this work affords some tolerance
to operating near the margins at high momentum by extrapolating away 
discretisation errors. The step
scaling proposed in this work is intended to raise the momentum
scale into a more convergent perturbative regime.

The structure of the paper is as follows.
In section~\ref{sec:background} we summarise the method of off-shell
non-perturbative renormalisation. In section~\ref{sec:formulation} we
introduce twisted boundary conditions for off-shell renormalisation.
This is motivated in section~\ref{sec:fourierbad} where we
explain how this solves a serious theoretical problem with defining the
continuum limit of operators renormalised using RI/MOM.
The implementation of twisted boundary conditions in the valence
sector for fermion vertex functions is discussed in 
section~\ref{sec:twistednpr}, and we
define lattice kinematics (\ref{sec:kinematics})
with which we intend to take the continuum
limit (\ref{sec:context}) for both exceptional and non-exceptional momenta. 
Our proposed step scaling scheme, based on these
techniques, is discussed in section~\ref{sec:stepscaling}.

We present data for the amputated vertex functions at a single
lattice spacing in section~\ref{sec:data}.
In section~\ref{sec:stepdata} 
we demonstrate practicality using existing configurations to take the
continuum limit for the first step in the step scaling programme.

\section{Background}

\label{sec:background}

The Rome-Southampton RI/MOM approach \cite{Martinelli:1994ty} 
involves a simple physically defined
momentum scheme, albeit with particularly 
unphysical renormalisation conditions. 
For any given operator, ${\cal{O}}$, a momentum configuration 
for the vertex function of that operator 
is chosen with
momenta $p_\mu$ selected from the Fourier modes for the simulated
lattice size $a L_\mu$, where $L_\mu$ is an integer:
$$a p_\mu = \frac{2 \pi}{L_\mu} n_\mu$$, with $n_\mu\in \{ 0,\ldots L_\mu \}$.

The renormalisation condition 
for any regularization scheme $S$ chooses counter terms such that
renormalised amputated vertex function, 
in Landau gauge and for a chosen set of external momenta,
is equal to the tree level operator. For example,
for bilinears this is
\begin{equation}
\Lambda^{S,ren}_{\cal O}(p_1,p_2)= \frac{Z_{\cal O}^S}{Z_q} \Lambda^S_{\cal O}(p_1,p_2) = {\cal O}^{\rm tree}, 
\end{equation}
where $\Lambda$ represents an amputated vertex function.
The original paper used the amplitude with
\begin{equation}
\label{eq:rimom}
p_1 = p_2 ; q^2 = 0.
\end{equation}
The approach facilitates the conversion between different schemes
because all that is required in any given scheme $S$ is a self-consistent
calculation of the relevant scattering amplitude to some order. 
By virtue of the physical definition, one scheme can be a 
fully non-perturbative lattice calculation with some 
lattice action at non-zero lattice spacing.
Thus, within the scaling window, Eq~(\ref{eq:window}), a lattice simulation can
be matched directly to continuum perturbation theory without use
of ill-convergent lattice perturbation theory.

This momentum window argument 
motivates the neglect of discretisation errors in a certain limit.
We shall see in section~\ref{sec:formulation}
that a rigorous continuum extrapolation using the original method 
is made impossible due to the use of Fourier modes. 
This is because keeping both the direction
and magnitude of the physical momenta fixed while changing the 
lattice spacing is incompatible with the discrete momentum spectrum.

\subsection{Non-exceptional momentum}

In the original RI/MOM kinematic choice a single gluon can carry all 
the hard momentum through the vertex leading to
infra-red effects  which fall only as $\frac{1}{p^2}$.
Depending on the operator these effects have a non-trivial 
dependence on the valence quark mass that can complicate mass
extrapolations, and in some cases require pion pole subtraction
\cite{Martinelli:1994ty,Blum:2001sr,Politzer:1976tv,Pascual:1981jr}.

RBC and UKQCD have developed the non-exceptional
momentum, SMOM, kinematic point as a preferred matching condition 
\cite{Aoki:2007xm}.
A significant gain comes from the extra suppression 
of non-perturbative effects. The perturbative expansion
must be calculated for non-exceptional momenta; this has been 
performed to one loop 
for bilinear operators \cite{Sturm:2009kb} and for $B_K$ \cite{IWBKcontinuum}.
Two recent publications have extended this to two loops for two different
schemes for the quark mass \cite{Gorbahn:2010bf,Almeida:2010ns}.
The non-exceptional kinematic point for bilinears has
\begin{equation}
\label{NEcond}
p_1^2 = p_2^2 = (p_1 - p_2)^2.
\end{equation}

RBC and UKQCD have 
found \cite{Aoki:2007xm} that both momentum and mass dependence are simplified
with non-exceptional kinematics. 
With this momentum flow, multiple hard gluons are required to 
create a soft sub-graph and non-perturbative infra-red effects,
such as spontaneous chiral symmetry breaking, fall as a higher power of the external momentum,
$\frac{1}{p^6}$ \cite{Yasumichi}. As our aim is 
to carefully study vertex functions while reducing the volume, we require
a momentum scheme that is not sensitive to condensate physics.

\subsection{Volume source NPR}

The original method used a single point source to calculate momentum space
Green's functions. 
The volume source technique was developed by QCDSF \cite{Gockeler:1998ye}. 
The attraction here is to evaluate
the amputated vertex function with the operator insertion averaged over
all $L^4$ lattice sites. 
%
We shall see it is easily possible to obtain 0.1\% statistical errors with volume
source techniques, and even smaller if hundreds of configurations or larger
volumes were used. With this statistical precision systematic effects like
${\cal{O}}(4)$ breaking lattice artefacts are visible and are in fact 
the dominant systematic error. These must either be 
included in the error analysis or better yet addressed using the techniques
of this paper to escape the Fourier constraints.

We use $i$, $j$ to represent colour indices, and $\alpha$, $\beta$ to represent 
spin indices. We define the four momentum source, used on a 
Landau gauge fixed configuration, as

\begin{equation}
\eta_p(x) = e^{i p_\mu x^\mu} \delta_{ij} \delta_{\alpha\beta},
\end{equation}
where the (dimensionless) momenta are $a p_\mu = n_\mu \frac{ 2\pi}{L_\mu}$.
On a given gauge field $U_\mu(x)$ we solve 
\begin{equation}
\sum_{x} D_{\rm dwf}(y,x) G(x,p) = \eta_p(y),
\end{equation}
and $D_{\rm dwf}$ is the Domain Wall fermion matrix \cite{Furman:1994ky}.
One propagator inversion for each leg momentum is
necessary, but this cost is more than offset by the gain in 
statistical accuracy from the volume average.

To compute the external legs required for amputating the 
vertex functions we require the momentum space
propagator
\begin{equation}
G(p_1,p_1) = \sum\limits_x e^{-i p_1^\mu x_\mu} G(x,p_1).
\end{equation}
We form phased propagators for each momentum:
\begin{equation} 
G^\prime(x,p) = G(x,p) e^{-ip\cdot x}
                  = \sum_{y} D_{\rm dwf}^{-1}(x,y)e^{ip\cdot (y-x)}.
\end{equation}
We select two independent momenta $p_1$, and $p_2$, and
form unamputated bilinear vertex functions $V_\Gamma$
for Dirac structure $\Gamma$:
\begin{equation}
V_\Gamma^{\text{bilinear}}(p_1,p_2)= \left[ \sum_x 
\gamma_5 (G^\prime(x,p_1))^\dagger \gamma_5
\Gamma 
G^\prime(x,p_2)\right]_{ij,\alpha\beta} ,
\end{equation}
and also for four quark operators: 
\begin{equation}
V_{\Gamma\Gamma}^{\text{4q}}(p_1,p_2) =\sum_x \left( 
\gamma_5 (G^\prime(x,p_1))^\dagger \gamma_5
\Gamma 
G^\prime(x,p_2) 
\right)_{ij,\alpha\beta}
 \left( 
\gamma_5 (G^\prime(x,p_1))^\dagger \gamma_5
\Gamma 
G^\prime(x,p_2) 
\right)_{kl,\gamma\delta}.
\end{equation}
Here, external colour and spin indices are left free
for off line amputation and projection.
This allows us to define the amputated Green's function as, for example,
\begin{equation}
\Pi_\Gamma(p) = \left( G^{-1}(p_1,p_1) V_\Gamma(p_1,p_2) \gamma_5 [G^{-1}(p_2,p_2)]^\dagger \gamma_5 \right)
\end{equation}
where $p^2 = p_1^2 = p_2^2 = (p_1 - p_2)^2$, and we take $q = p_1-p_2$.
The bare vertex amplitudes are then obtained by projecting the amputated 
Green's functions onto their tree level values, for example
\begin{equation}
\Lambda_\Gamma = \frac{1}{12} Tr \left( \Pi_\Gamma \Gamma \right).
\end{equation}

\section{Controlled continuum extrapolation}
\label{sec:formulation}

We will motivate and introduce the use of
twisted boundary conditions in the context of off-shell renormalisation.
This will enable us to develop a framework for better controlled
continuum extrapolation than is possible using
Fourier modes.

\subsection{Continuum extrapolation and Fourier constraints}
\label{sec:fourierbad}
In continuum Euclidean space simultaneously
rotating all momenta in a scattering amplitude by any
$O(4)$ matrix must give equivalent results.
However, in our discrete system there is only $H(4)$ symmetry and, 
even in the infinite volume, only momenta related by $\pi/2$ rotations and 
reflections are equivalent. Vertex functions will receive \emph{different}
$O(a^2)$ errors depending on the direction of the momentum relative
to the lattice axes. The original studies
did not resolve ${\cal O}(4)$ breaking
lattice artefacts due to statistical imprecision. Different
${\cal O}(4)$ equivalent but ${\cal H}(4)$ distinct 
momentum configurations, given by different Fourier modes,
were treated interchangeably and simply averaged together. This is
backed to some extent by the Rome-Southampton window argument for a
region of safe operation with respect to lattice artefacts. 

When the method outlined above is combined with
a continuum extrapolation, the naive use of ${\cal H}(4)$ inequivalent
momenta as interchangeable is theoretically unsound because
$O(4)$ breaking is a component of the lattice artefacts that 
should be removed by the extrapolation.
If the orientation of momenta relative to lattice axes does not remain fixed
as $a$ is changed, 
an extrapolation in $a$ is \emph{not} formally valid as one uses 
a \emph{different} observable for each lattice spacing 
(differing due to ${\cal H}(4)$ inequivalent momentum choices).
The Fourier constraints make it difficult to select 
the same physical momentum simultaneously on more than one 
lattice spacing, and these differing $O(4)$ breaking lattice artefacts
necessarily enter each data point in 
a continuum extrapolation of data renormalised using the Rome-Southampton
method. This introduces an intrinsic systematic uncertainty at a level set
by the size of $O(4)$ breaking artefacts. 

$O(4)$ breaking effects were argued to be ignorable as they were not
statistically resolved.
This justifies smooth interpolation in $p^2$
to obtain matched physical momenta. 

It also somewhat justifies an (otherwise dubious) continuum extrapolation. 
We measure the vertex function observable on different lattice spacings
at the same physical momentum magnitude, but must pick ${\cal H}(4)$
inequivalent  momenta to satisfy  Fourier constraints. This results
in incorrectly parametrised lattice artefacts, 
however ${\cal O}(4)$ breaking effects can be safely ignored when they 
are substantially smaller than statistical errors.

With the volume source technique, however,  $O(4)$ breaking effects are well
resolved and are a dominant systematic error. This compromises our
ability to perform a continuum extrapolation, and
these considerations 
represent a serious problem that we address in this paper with the use
of twisted boundary conditions. 

\subsection{Twisted boundary conditions}
\label{sec:twistednpr}

From the discussion of the previous subsection~\ref{sec:fourierbad} 
we see that
it would be rather better to crisply remove lattice artefacts 
by continuum extrapolation, and only apply perturbative matching in
the continuum limit. 
This is difficult when constrained to use Fourier modes.
Applying twisted boundary conditions
to propagator inversions enables arbitrary momenta
to be used. The twisted boundary approach
differs by only finite volume effects
from a simulation on a much larger lattice
with an exact Fourier mode of the same momentum.

In order to simultaneously satisfy the constraint \eqref{NEcond} and only 
use one momentum direction we propose choosing one kinematic satisfying
\eqref{NEcond} and use twisted boundary conditions 
\cite{Boyle:2003ui,Bedaque:2004kc,deDivitiis:2004kq,Flynn:2005in,Boyle:2007wg},
to vary the magnitude of the momentum. 
The twisting technique has been used to insert arbitrary three-momenta
in form factor calculations with the success demonstrated, for example, by the direct
comparison of refs \cite{Boyle:2007qe,Antonio:2007mh} to refs 
\cite{Boyle:2007wg,Boyle:2008yd,Boyle:2010bh}.
which make use of the same configurations and action.

In this paper the technique is used to allow arbitrary four-momenta to be used
\cite{Boyle:2003ui}, and hence
allow to rigorously disentangle discretisation and perturbative 
truncation effects for the first time. Of course, the lattice artefacts
are still present at finite lattice spacing but, up to finite volume effects,
these become the \emph{same} lattice artefact at each lattice spacing.
This is not the case if different Fourier modes are used.
The goal is to enable us to determine the non-perturbative anomalous running
in the RI scheme in the continuum limit prior to the use of continuum
perturbation theory. 

We now formulate the momentum space propagator calculation with 
twisted boundary conditions for use in off-shell renormalisation.
Let the quark fields satisfy twisted boundary conditions
$q(x+L) = e^{i\theta}q(x)$ and define \cite{Sachrajda:2004mi}
\begin{equation}
\tilde{q}(x) = e^{-iBx}q(x)
\end{equation}
with $a B_\mu = \frac{\theta}{L_\mu}$  so 
that $\tilde{q}(x)$ satisfies periodic boundary conditions. 
This transformation changes the continuum Dirac operator:
\begin{equation}\label{dirac}
D = (\cancel{\partial} + m) \rightarrow \tilde{D} = (\cancel{\partial} + i\cancel{B} + m)
\end{equation}
$\tilde{D}$ has inverse $\tilde{G}$ and $D$ has inverse $G$; they are related by
\begin{equation}\label{twprop}
G(x,y) = e^{iB(x-y)} \tilde{G}(x,y)
\end{equation}
using translational invariance which we should recover after gauge averaging.
Let 
\begin{equation}
\tilde{G}(z,p) =  \sum_x \tilde{G}(z,x)e^{ipx}
\end{equation}
then
\begin{equation}
\sum_z \tilde{D}(y,z)\tilde{G}(z,p) = e^{ipy}
\end{equation}
So inverting the twisted Dirac operator \eqref{dirac} with a momentum source gives $\tilde{G}(z,p)$. Note that $\tilde{G}(z,x)$ satisfies periodic boundary conditions so that in the source term, $e^{ipy}$, 
$a p_\mu = \frac{2 \pi n_\mu}{L_\mu}$ is a Fourier mode.
$\tilde{G}(z,p)$ is related to $G(z,p)$ via,
\begin{equation}\label{Gtilde}
\tilde{G}(z,p) = \sum_x e^{-iB(z-x)} G(z,x) e^{ipx} = e^{-iBz} G(z,p+B).
\end{equation}
The propagator can be obtained from,
\begin{equation}
\tilde{G}(p,p) = \sum_z e^{-ipz} \tilde{G}(z,p) = \sum_{z,x} e^{-i(p+B)(z-x)} G(z,x) = G(p+B,p+B)
\end{equation}
Thus the net effect of twisted boundary conditions is to shift the momentum $p$  to $p + B$ where B is arbitrary. In order to compute the non-exceptional Green's functions observe,
\begin{equation}\label{Gp1p2def}
 \tilde{V}_{\Gamma}(p_1,p_2) = \sum_{x,y,z} e^{-i p_1 (x-z)}\tilde{G}(x,z) \Gamma e^{-i p_2 (z-y)}\tilde{G}(z,y) = \sum_{z} \gamma_5 e^{i p_1 z} \tilde{G}(z,p_1)^\dagger \gamma_5  \Gamma e^{-i p_2 z} \tilde{G}(z,p_2)
\end{equation}
Using the inverse of \eqref{Gtilde} $\tilde{G}_{\Gamma}(p_1,p_2)$ the
vertex function is seen to be,
\begin{equation}\label{Gp1p2defa}
 \tilde{V}_{\Gamma}(p_1,p_2) = \sum_{x,y} \gamma_5 e^{i (p_1+B_1)x}G(x,p_1+B_1)^\dagger \gamma_5 \Gamma e^{-i (p_2+B_2)y}G(y,p_2+B_2) = V_{\Gamma}(p_1+B_1, p_2+B_2)
\end{equation}

We consider the values for $p_1$, $p_2$, $B_1$ and $B_2$ in the following
subsection.

\subsection{Kinematics}
\label{sec:kinematics}

\begin{figure}[htb]     
        \begin{center}
\includegraphics*[width=0.5\textwidth]{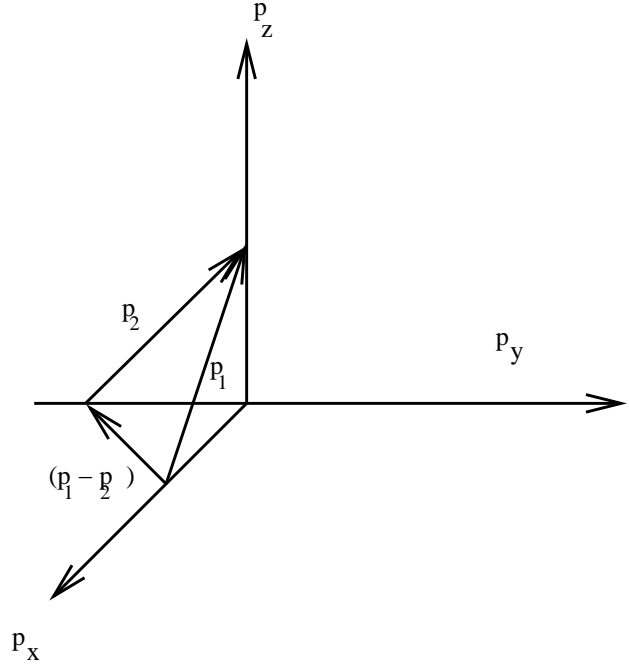}
\end{center}
\caption{
Non-exceptional momentum configuration used in this paper. 
The momenta $p_1$, $p_2$ and $(p_1-p_2)$ must be equal in magnitude
and are represented as an equilateral triangle with vertices
touching the $p_x$, $p_y$, and $p_z$ axes. With twisted boundary conditions
the sides of this triangle can be continuously scaled allowing
both smooth interpolation of momentum dependence and controlled continuum
extrapolation. This is very much in contrast to the situation
that arises when using only Fourier modes.
\label{fig:mom}
}
\end{figure}
We select Euclidean momenta in the direction of $p_1 = (-1,0,1,0)$ and $p_2 = (0,1,1,0)$ up 
to $H(4)$ symmetry operations. 
This choice minimises $S^4 = \sum_{i} p_i^4$ which we
take as measure of discretization errors, subject to the constraint
that $p_1$/$p_2$/$(p_1-p_2)$ be $H(4)$ equivalent momenta.
It is certainly possible to rotate some combination of $p_x,p_y,p_z$
into the temporal direction breaking this $H(4)$ equivalence, and this would
form an interesting possibility to demonstrate universality of the continuum
limit in a later work. 

The non-exceptional $p_1$/$p_2$/$(p_1-p_2)$ 
kinematics can be represented by an equilateral triangle 
in momentum space with vertices 
lying on the $p_x,p_y,p_z$ coordinate axes (figure~\ref{fig:mom}). 
Continuous dilation of this triangle 
can be performed by using twisted boundary
conditions with a fixed orientation of the 
$p_1$ , $p_2$, $p_1 - p_2$ triangle. This continuity
allows one to pick a fixed MOM observable as the
lattice spacing is varied.
In particular, we choose parallel twisting vectors 
in the directions
$B_1 = (-\theta, 0, \theta, 0)$ corresponding to $p_1$ and 
$B_2 = (0, \theta, \theta, 0)$ corresponding to $p_2$. We
continuously vary $\theta$ to move the vertices of the triangle 
along the axes.

The non-exceptional momenta are chosen to have small discretisation 
errors by spreading the power across multiple coordinate axes. 
In general for any non-exceptional configuration there will be $O(a^2)$ errors.
For exceptional momentum amplitudes we take vertex functions with $p_1$
on both legs.
As the direction is kept fixed the vertex function is a smooth 
function of $(ap)^2$ and interpolation in momentum is possible. 
Extrapolation to the continuum limit at fixed physical momentum
is now theoretically justified.

\subsection{Continuum extrapolation}
\label{sec:context}
Between different lattice spacings we compare only
renormalised quantities and their ratios.
This is quite natural as we desire to take the continuum limit of the 
product of a bare operator matrix element and its RI scheme 
renormalisation constant at some accessible scale $p$. 
The scaling factors for each operator to match between different 
$\beta$'s introduced in \cite{Zhestkov:2001hx} are then not required.
The aim of the latter part of this
paper will be to raise, in the continuum limit, this 
scale $p$ to one ($s^n p$ where $s$ is a scale factor)
where there is better perturbative convergence.

We eliminate $Z_q$ via the vertex function 
of the (conserved) vector or axial currents:
\begin{equation}
R_{\cal O}(p,a,m) = \frac{\Lambda_A(p,a,m)}{\Lambda_{\cal O}(p,a,m)} = \frac{Z_{\cal O}(p,a,m)}{Z_A}
\end{equation}
and this ratio will be extrapolated to the chiral
limit to produce a renormalization constant
\begin{equation}
Z_{\cal O}(a,p) = Z_A \lim_{m\to 0} R_{\cal O}(p,a,m)
\end{equation}
where, for convenience, we use the vertex function of the 
local rather than conserved axial current and eliminate this
with a previously computed $Z_A$.
We then continuum extrapolate the product of $Z_{\cal O}$ with $\langle {\cal O} \rangle$:
\begin{equation}
\langle {\cal O} \rangle^{\rm ren} = \lim_{a\to 0}\langle {\cal O} \rangle Z_{{\cal O}}.
\end{equation}
This gives $\langle \cal O \rangle $ in the RI/MOM scheme at a scale determined by the physical momentum chosen which must be the same scale on all the lattices used to take the continuum limit.

We also consider the factor required 
to convert $\langle {\cal O} \rangle$  at scale $p$ to 
scale $sp$ where $s>1$ is a scale factor.
This is
\begin{equation}
\Sigma_{\cal O}(p,s p,a) = \lim_{m\to 0}
                        \frac{R_{\cal O}(sp,a,m)}{
                             R_{\cal O}(p,a,m)},
\end{equation}
and its continuum limit is
\begin{equation}\label{eq:sig}
\sigma_{\cal O}(p,s p) = \lim_{a\to 0}\Sigma_{\cal O}(p,s p, a).
\end{equation}
$\sigma_{\cal O}(p,s p)$  
encodes the scale dependence of the RI scheme 
in the continuum limit expanded around the scale $p$.
We can then assess the degree to which
the running around our match scale is  
perturbative without risk of confusion by
possible lattice artefacts.
By picking a well defined momentum orientation for the
amplitude as a function of $a$, we remove the ambiguity in 
taking the continuum limit that arises
when selecting different Fourier modes for each value of $\beta$.

We have demonstrated how to 
obtain a controlled expansion in powers of the lattice spacing
for amputated vertex functions and gain better
control of lattice artefacts via continuum extrapolation. 

\section{Step scaling}

\label{sec:stepscaling}

QCD perturbation theory at lattice scales is not rapidly convergent, 
and a means to increase this scale without applying brute force to raise the 
lattice cut-off is required.
Step-scaling \cite{Luscher:1991wu,Luscher:1993gh}
is the natural approach to do this where, in a
series of simulations, the physical volume is reduced as the 
lattice spacing is reduced to enable the renormalisation
scale to be raised without the cost associated with simulating
all scales on a single lattice.

\subsection{Step scaling background}

Step scaling has been well developed in the Schr\"odinger
functional approach 
\cite{Luscher:1991wu,Luscher:1993gh,Bode:2001jv,Sint:1995rb,Sint:1994en,Luscher:1992zx,Luscher:1992an,Bode:2001jv}. 
A finite volume scheme based on Creutz ratios has also been recently
developed \cite{Bilgici:2009kh}.
In this paper we seek to develop
a related approach based on the Landau gauge fixed RI/MOM method. 
Our approach does not tie the renormalisation 
scale to the volume, and therefore also avoids the need to fine tune 
bare couplings to precisely match a volume sensitive 
renormalised coupling between simulations.

After reviewing existing step scaling methods we will
attempt to address perturbative errors associated with the
low-end of the Rome-Southampton scaling window by raising the scale
at which we match to perturbation theory by introducing a new
step scaling scheme. In order to use the same notational conventions as the 
literature on step scaling in the Schr\"odinger functional scheme, for the 
rest of this section 
let $L$ represent the length of the lattice in physical units rather 
than units of lattice spacing.
This departs from our convention elsewhere in the
paper, which takes $L$ as an integer
lattice size and 
is consistent with notation typically used for RI/MOM renormalization.

\subsubsection{Schr\"odinger functional}

In step scaling there is typically a renormalisation condition 
imposed that is physically defined in a fictional finite volume universe.
In the case of the Schr\"odinger functional this is the 
coupling $\bar{g}(\mu)$ at a scale $\mu=\frac{1}{L}$ 
defined by the volume. A second simulation, at the same 
bare coupling and with larger volume $s L$, then gives the coupling at
a scale $\mu^\prime = \frac{\mu}{s} = \frac{1}{s L}$
$$\bar{g}(\mu^\prime ,a) = \Sigma(s,\bar{g}(\frac{1}{L}),\frac{a}{L}).$$ 
Here $s>1$ is a scale factor, and is typically $s=2$.
The continuum limit $$\sigma(s,\bar{g}(\frac{1}{L})) = \lim_{a\to 0}\Sigma(s,\bar{g}(\frac{1}{L}),\frac{a}{L})=\lim_{a\to 0}\bar{g}(\mu^\prime = \frac{1}{s L},a),$$ 
is  taken \emph{while holding the measured coupling $\bar{g}(\mu=\frac{1}{L})$
fixed}. Hence $\bar{g}(\mu=\frac{1}{L})$ has no $a$ dependence
and the trajectory to the continuum limit leaves the
renormalised coupling $\bar{g}(\mu=\frac{1}{L})$ lattice artefact free at 
(and only at) the scale $\mu=\frac{1}{L}$.

However, $\bar{g}(\mu^\prime ,a)$ is lattice spacing dependent, with
discretisation errors that 
are removed by taking this continuum limit.
Thus
$\bar{g}(\mu=\frac{1}{L})$ is the
observable quantity used to define the lattice spacing and hence physical
volume. The Sch\"odinger functional is particularly economical since
this is in fact the most infra-red scale accessible within the
simulated volume.

When two consecutive scale evolution steps are considered,
two sequences of simulations must be performed to determine the 
evolution of the coupling from  scale $\mu^\prime_1 \to \mu_1$, 
and from $\mu^\prime_2 \to \mu_2$ in such a 
way that $\mu_2 = \mu_1^\prime$. Since \emph{different quantities}
$\bar{g}(\mu_1)$ and $\bar{g}(\mu_2)$ are held fixed
for the two sequences of simulations (and effectively
determine the lattice spacings), satisfaction of the 
constraint $\mu_2 = \mu_1^\prime$ in the continuum limit
is ensured by defining the scaling trajectory for the second
sequence such that $\bar{g}(\mu_2,a) =  \bar{g}(\mu_1^\prime,a\to 0)$.

Since each simulation must correctly describe the length scales associated 
with the quantity used to determine trajectory to the continuum limit,
carefully changing this between steps was a key component 
allowing the volume to be reduced.
This is a feature we intend to reproduce in our method.

In the Schr\"odinger functional $\bar{g}(\frac{1}{L})$
defines the scaling trajectory and is \emph{directly} coupled to the volume. 
Taking the continuum limit holding  $\bar{g}(\frac{1}{L})$ fixed
then requires fine tuning of $\beta$ to \emph{exactly} match
the desired non-perturbative coupling -- and hence to match 
physical volumes defined by using $\bar{g}(\frac{1}{L})$ to set the scale.
This is a fine tuning step that our proposal below avoids. 

\subsubsection{Previous work on step scaling RI/MOM}

The possibility of step scaling with RI/MOM has been previously studied
\cite{Zhestkov:2001hu}, \cite{Zhestkov:2001hx}. 
This work used a series of quenched configurations 
where the ratio of lattice spacings had been 
tuned to be precisely a factor of two to obtain aligned Fourier
modes. This fine tuning is expensive in a dynamical simulation.
Use of power counting and the Rome-Southampton scaling
window was made, rather than the controlled continuum extrapolation at
fixed physical momentum introduced in this paper. 
Free parameters for each operator 
were introduced to match the renormalisation 
constants for different $\beta$'s 
corresponding to the (possibly non-perturbative) 
anomalous scaling of the operator with the lattice cut-off. This excellent 
start was not easily developed into a practical technique.

\subsection{RI/MOM Step Scaling Proposal}

In order to develop a practical step scaling scheme we require several pieces.
Firstly, we have defined continuum limit non-perturbative evolution
ratios that can be used to access higher momentum scales. 
Step scaling will combine a sequence of these objects,
each determined inexpensively compared with a brute force method.

We believe that three technical advantages we have over the earlier attempt
at RI/MOM step scaling make the task more tractable.
The first, volume sources, reduces statistical errors greatly giving precision to the approach. The second, non-exceptional momenta, renders mass
dependence and infra-red behaviour in $p^2$ benign; this 
greatly assists with both chiral extrapolation and matching between ensembles,
in addition to finite volume sensitivity. These were described above.

In this paper we have introduced a third advantage:
using twisted boundary conditions to select arbitrary four momenta.
The direction of the scattering momenta
can be kept fixed relative to lattice axes and
arbitrary values of $p^2$ can be chosen. This is a Good Thing because
it allows the same physical momenta according to the lattice
symmetries $H(4)$ to be chosen on each ensemble. The vertex
function observable will \emph{only then} have a valid expansion in powers
of the lattice spacing $a$.
This approach enables precise matching of momenta between ensembles, 
and precise continuum extrapolation.

We will keep the off-shell momentum scales hard enough that the physical
volume should not be resolved. In this way the the calculation will be
finite volume effect safe, and the perturbative matching will 
use the standard infinite volume perturbation
theory already available to high order for off-shell renormalisation.
These are significant advantages over the Schr\"odinger functional.

For the scheme to be affordable it is therefore necessary that it
be able to operate in a small physical volume. 
For offshell amplitudes this is the case whenever the
virtuality is too hard to resolve the finite volume 
\begin{equation}
p^2 \gg (\frac{\pi}{L})^2.
\end{equation} 
Thus, for the purposes of step scaling, the window condition
Eq~(\ref{eq:window}) becomes 
\begin{equation}
\label{eq:sswindow}
(\frac{\pi}{L})^2 \ll p^2 \ll \left(\frac{\pi}{a}\right)^2.
\end{equation} 
This latter condition is likely possible to meet using
modest lattice volumes $\frac{L}{a} \le 16$ 
at all stages of the calculation.
We will compute vertex functions
on lattices at different values of $\beta$ with 
overlapping scaling windows. This will then enable the determination of
a continuum step scaling function
for some operator ${\cal O}$ at a physical scale $p^2$ chosen in the window 
Eq~(\ref{eq:sswindow}) to that at a scale larger by a scale factor $s$.
This is precisely $\sigma_{\cal O}(p, s p)$ from the previous section.

These can be combined in a series of non-perturbative steps to a high
scale, with perturbative conversion to $\overline{MS}$ is then applied
only where it is well convergent. For example,
\begin{equation}
\begin{array}{ccc}
\langle {\cal O}^{\overline{MS}}(\mu) \rangle
&=&
\langle {\cal O}^{\rm SMOM}(p) \rangle\\
&\times&
\sigma^1_{\cal O}(p,s p)\\
&\times&
\sigma^2_{\cal O}(sp,s^2p)\\
&\ldots&\\
&\times&
\sigma^n_{\cal O}(s^{n-1}p,s^np)\\
&\times&
\left[1 + c^{\rm SMOM\to\overline{MS}}\alpha_s(\mu = s^n p)\right].
\end{array}
\end{equation}
It appears clear that Eq~(\ref{eq:sswindow}) can be satisfied at reasonable
expense. For example $16^3$ domain wall fermion simulations will be 
inexpensive with the next generation of supercomputers; multiple such
ensembles dedicated solely to the renormalisation of lattice operators
are quite affordable. This is particularly helped by the relatively
benign mass dependence of non-exceptional momentum vertex functions.
It is also clear that on these proposed small ensembles all hadronic
quantities must be avoided to ensure finite volume safety of the analysis.
For the determination of $m_{\rm res}$, the PCAC is an operator relation and
so it is immaterial whether a finite volume or a physical pionic state
is used to determine the ratio. Determining the lattice spacing
(or more specifically the ratio of the lattice spacing to that of an earlier
larger volume simulation) in a finite volume safe manner is discussed below.

\subsubsection{Trajectory to continuum limit}

\label{sec:traj}

We now define how the continuum limit of
$\Sigma^n_{\cal O}(s^{n-1} p,s^n p,a)$ 
should be taken to obtain
$\sigma^n_{\cal O}(s^{n-1}p,s^n p)$. 
In order to avoid the fine tuning problem
we seek a quantity $q$ associated with a continuously variable length 
scale $L_q$ significantly shorter than the lattice extent $L$  to 
determine the trajectory to the continuum limit.
We plan to reduce the volume successively with each step,
and a different definition of the
lattice spacing (e.g. a different $L_q$) must be used for each step.

In the Schr\"odinger functional method this is done by fine tuning 
(i.e. constraining) 
$\bar g(\mu_{n+1})$  for each non-zero lattice spacing entering the continuum extrapolation
of $\sigma^{n+1}$ to match the continuum limit of the high scale coupling
obtained from the previous step, $\lim_{a\to 0}\bar g(\mu^\prime_n,a)$. 
We seek a the family of scale determining 
quantities $\{q_n\}$ that each be used to determine the 
lattice spacing in the corresponding $n$-th step of the step scaling scheme.
These will play part of the role that $\bar g(\mu=\frac{1}{L})$ plays for
the Schr\"odinger functional.


The continuum limit of successive quantities $q_n$ in the sequence 
will be determined non-perturbatively as we proceed. Previously
determined values will be used to define 
the trajectory to the continuum limit for successive steps.
As a concrete example our initial results we will take 
$q_n$ to be the static inter-quark force and so the determination of
the lattice spacings corresponds to a family of Sommer
scales based on the static potential. In the Schr\"odinger functional
the quantity $q_n$ corresponds to the coupling $\bar g(\frac{1}{s^n L})$.

At step $n$ we will determine the continuum limit of $q_{n+1}$ while
determining the lattice spacing $a^{(q_n)}$ from $q_{n}$, and thus
must maintain the constraint $q_n(a) = q_n^{\rm cont}$ while taking the
continuum limit:
\begin{eqnarray}
 q_{n+1}^{\rm cont} &= &
\lim_{a^{(q_n)}\to 0} \left. q_{n+1}(a^{(q_n)})\right|_{q_{n}(a) = q_n^{\rm cont}},
\end{eqnarray}
so that, for example, we determine the continuum limit of the inter-quark
force at a reduced length scale in one step, and then reuse this to constrain
the trajectory to the continuum limit in the next step. Using this trajectory
we define the scale evolution functions $\sigma^n$,
\begin{eqnarray}
\sigma^n_{\cal O}(s^{n-1}p,s^np) &=& \left. \lim_{a^{(q_n)}\to 0} \Sigma^n_{\cal O}(s^{n-1}p,s^np,a^{(q_n)}) 
\right|_{
q_n(a) = q_n^{\rm cont}},
\end{eqnarray}
where $q_n^{\rm cont}$ is determined by the previous step. 

When $q$ is a function of a continuous scale $L_{q} \ll L$, this distance
can be varied post simulation and without finite volume effects, and this allows us to avoid 
fine tuning $\beta$.
Ultimately this process will become difficult as all dependence on lattice
spacing becomes logarithmic and precision quantities are
required; however, there may also be less appropriate but expedient
choices that enable immediate progress for a few steps 
away from our relatively coarse simulations.
We consider two possibilities.

\subsubsection*{Static potential}

We note the static potential has been measured successfully over a large
range of length scales in the quenched approximation with the Wilson
gauge action \cite{Necco:2001xg}. This involved the use of
a shorter length scale $r_c\simeq 0.26 {\rm fm}$ than the more 
common $r_0 \simeq 0.48 {\rm fm}$  
\cite{Sommer:1993ce}. 
This calculation is also interesting because it 
successfully fits $\log \frac{a}{r_0}$ as a polynomial
in $\beta$. Such an approach may ultimately assist continuing 
lattice spacing determinations to increasingly fine and small volume 
simulations.

We consider a sequence of scales, of the same class as the Sommer
scale 
\begin{equation}
r_n^2 F(r_n) = C_n.
\end{equation}
The Sommer scale $r_0$ takes 
\begin{equation}C_0 = 1.65\end{equation}

Thus a step scaling scheme with scale factor $s$ can then be defined 
choosing $p_n = s^n p$ and $r_n = \frac{r_0}{s^n}$ as follows:

\begin{itemize}
\item Determine $\sigma(p_0, p_1)$ in continuum limit holding $r_0 p_0$ fixed such that $r_0^2 F(r_0) = C_0$
\item Determine $C_1 = \frac{r_0^2}{s^2} F(\frac{r_0}{s})$ in continuum limit holding $r_0$ fixed
\item Decrease $L$ by $\simeq {\frac{1}{s}}$ without fine tuning
\item Determine $\sigma(p_1, p_2)$ in continuum limit holding $r_1 p_1$ fixed such that $r_1^2 F(r_1) = C_1$
\item Determine $C_2 = \frac{r_1^2}{s^2} F(\frac{r_1}{s})$ in continuum limit holding $r_1$ fixed
\item Decrease $L$ by $\simeq {\frac{1}{s}}$ without fine tuning\\
etc...
\end{itemize}

Following a rule of thumb that $r < \frac{L}{3}$ for the static
potential should ensure finite volume safety and enable simulation down to 
$L \simeq 0.75 {\rm fm}$ when using scales similar to $r_c$ to set the scale. 
Eventually this will become imprecise when entering a region
where the potential runs logarithmically. However
we believe a substantial benefit is already achievable prior
to addressing this issue.

\subsubsection*{Alternative schemes}

In principle, we can use the momentum dependence of the off-shell
vertex functions themselves to match the lattic scales between different
ensembles. {This may assisted by the improved techniques of this paper,
particularly the tunable momentum scale which
allows us to select the length scale that sets lattice spacing.
This will be the subject of further study.}

If we accept fine tuning, it is also possible to 
combine step scaling of off-shell vertex
functions with existing finite volume schemes such as the Schr\"odinger
functional or the Wilson loop scheme used to match lattice spacings.

\section{Results}

We use the domain wall fermion action \cite{Furman:1994ky}, and 
$16^3\times 32$ ensembles with $L_s=16$ and with the Iwasaki gauge action
\cite{Iwasaki:1985we}. We use two
ensembles sets in this paper, with $\beta = 2.13$ and 
$\beta = 2.23$. The $\beta=2.13$ ensembles
were studied extensively in Ref. \cite{Allton:2007hx}.
The three $\beta=2.13$ $16^3$ ensembles used in this section 
have strange quark mass $m_h=0.04$ and degenerate light quark masses $m_l=0.01,0.02,0.03$.
The second ensemble set with $\beta=2.23$ is not previously published 
and was used as part of a parameter search,
made in the style of \cite{Antonio:2006px,Antonio:2008zz}, prior to
our $32^3$ simulations \cite{thirtytwocubed} which lie nearby in parameter space.
There are two ensembles with $m_h=0.04$ and $m_l=0.01,0.02$.  There are around 2000 trajectories
in these ensembles.

For $\beta=2.13$ the extrapolations to $m_q = -m_{res}$ were made with $m_{\rm res} = 0.00305$,
$Z_A = 0.7161(1)$, and the lattice spacing was taken as $a^{-1} = 1.729$ GeV  \cite{Allton:2008pn}. 
We find in this paper that $\beta=2.23$ corresponds to an inverse
lattice spacing of around $a^{-1} \simeq 2.14$GeV, 
we estimate $m_{\rm res}\simeq 10^{-3}$ and $Z_A = 0.740$ by interpolating between nearby values
for $\beta$. This is not ideal, but adequate for the purposes of this
demonstration.

All of the vertex function data was computed using the volume source 
method with twisted boundary conditions as described in 
section \ref{sec:twistednpr} using 20 gauge configurations for 
each mass, see tables \ref{tab:gauge2.13} and 
\ref{tab:gauge2.23} with momenta in table \ref{tab:momenta}.

\begin{table}[hbt]
\begin{tabular}{ccc}
\hline
\hline
$a m_q$ & Range & $\Delta$ \\
\hline
0.03   & 800 -  2320  & 80\\
0.02   & 1000 - 2520  & 80\\
0.01   & 1000 - 2520  & 80\\
\hline
\end{tabular}
\caption{\label{tab:gauge2.13}
$\beta = 2.13$ lattice giving the range of molecular dynamics time and the separation $\Delta$ between gauge configurations used in this work. Because so few configurations are needed with the volume averaging a large $\Delta$ was chosen to minimise auto-correlation effects.
}
\end{table}

\begin{table}[hbt]
\begin{tabular}{ccc}
\hline
\hline
$a m_q$ & Range & $\Delta$ \\
\hline
0.02   & 1000 - 1760  & 40\\
0.01   & 1240 - 2000  & 40\\
\hline
\end{tabular}
\caption{\label{tab:gauge2.23}
$\beta = 2.23$ lattice giving the range of molecular dynamics time and the separation  $\Delta$ between gauge configurations used in this work.
}
\end{table}

\begin{table}[hbt]
\begin{tabular}{cccc}
\hline
\hline
$\beta$ & $a p_1$ & $a p_2$ & Range\\
\hline
2.13   & $(0,x,x,0)$ & $(-x,0,x,0)$  & $x=\frac{2 \pi}{L}(1.875) \rightarrow x=\frac{2 \pi}{L}(2.75)$ \\
2.23   & $(0,x,x,0)$ & $(-x,0,x,0)$  & $x=\frac{2 \pi}{L}(1.5625) \rightarrow x=\frac{2 \pi}{L}(3.125)$\\
\hline
\end{tabular}
\caption{\label{tab:momenta} 
The momenta at which we calculated the vertex functions. We note that the momentum $\frac{2 \pi}{L}(0,1.875,1.875,0)$, for example, can be reached using 'base' momentum $\frac{2 \pi}{L}(0,1,1,0)$ and twist $\frac{2 \pi}{L}(0,0.875,0.875,0)$ or base $\frac{2 \pi}{L}(0,2,2,0)$ and twist $\frac{2 \pi}{L}(0,-0.125,-0.125,0)$. We chose the base momentum such that the magnitude of the twisting component was less than $\frac{2 \pi}{L}(0.75)$, but such a choice is entirely arbitrary.
}
\end{table}

\subsection{Vertex functions with twisted boundary conditions at a single lattice spacing}
\label{sec:data}

\begin{figure}[htb]     
        \begin{center}
\includegraphics*[width=0.75\textwidth]{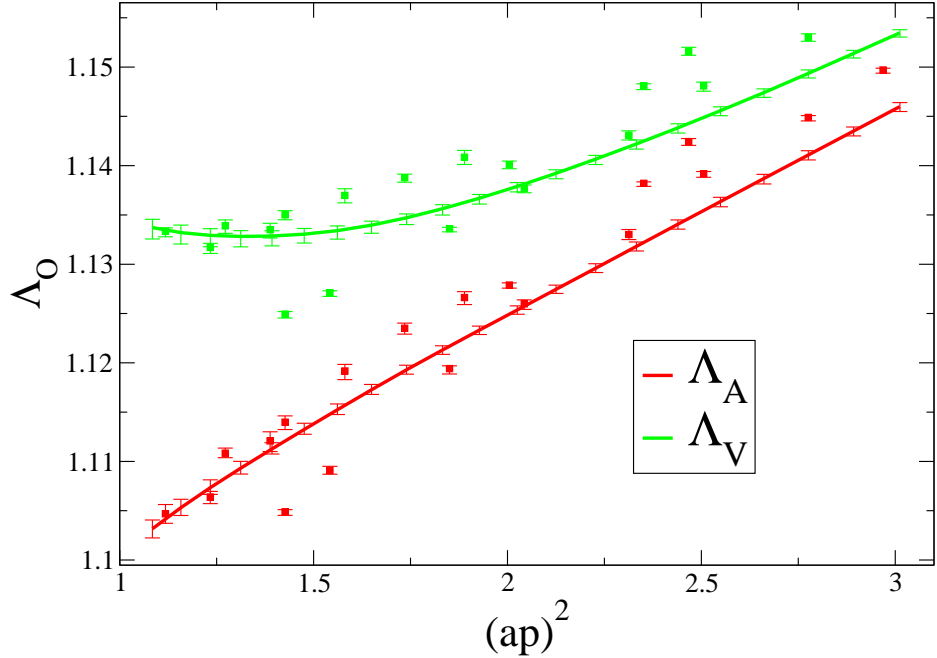}
\end{center}
\caption{\label{fig:cftw} 
The axial (red) and vector (green) vertex functions on the $\beta = 2.13$ lattice at $m_q = 0.03$, see Section \ref{sec:data} for more details.  Comparing untwisted (squares) with twisted (connected) data; twisting completely eliminates the $O(4)$ breaking scatter.
}
\end{figure}

In this section we focus on the data obtained using the $\beta=2.13$ ensembles and display
the quality of data obtained on a single lattice spacing.

Figure~\ref{fig:cftw} contains a comparison of the traditional volume averaged
Fourier mode NPR to our new twisted boundary condition technique. We show the 
vector and axial vertex functions at fixed quark mass, $m_q = 0.03$, and
with exceptional momentum kinematics on 
the $\beta = 2.13$ lattice, using 10 configurations.
$O(4)$ breaking lattice artefacts produce a large scatter in the data using the
traditional approach and this is scatter completely removed by 
the twisted boundary conditions technique. Of course, 
lattice artefacts are still present and of the same size,
but the key point is that we can now vary $\beta$ while looking at the 
same off-shell momentum in order to extrapolate these away in the continuum 
limit.

For the exceptional kinematic configuration we use the projectors 
of \cite{Aoki:2007xm} and 
the perturbative running and matching
calculated in \cite{Chetyrkin:1999pq} for mass and \cite{Gracey:2003yr} for the tensor current. 
For ${\cal O}_{VV+AA}$ we use the results of \cite{Buras:2000if,Ciuchini:1997bw}.
We denote these exceptional momentum schemes RIMOM.

For non-exceptional kinematics we compare the two schemes introduced in
\cite{Sturm:2009kb}. The first is the scheme of \cite{Sturm:2009kb} 
which corresponds to choosing projectors 
$\cancel{q} q_\mu$ for the vector vertex function and $\cancel{q} q_\mu \gamma_5$ for the axial 
vertex function. We denote this scheme SMOM-$\cancel{q}$.
The second uses $\gamma_\mu$ and $\gamma_\mu \gamma_5$ for vector and axial vertex functions, 
we refer to this scheme as SMOM-$\gamma_\mu$.
The one-loop matching and two-loop anomalous dimensions for tensor current and mass 
are given in \cite{Sturm:2009kb}. These results have recently been extended to two
loop matching and three loop anomalous dimensions \cite{Almeida:2010ns,Gorbahn:2010bf}.
For ${\cal O}_{VV+AA}$ we use several unpublished new schemes and perturbative
results by Sachrajda and Sturm \cite{IWBKcontinuum}, and we thank
them for their private communications.

\begin{figure}[htb]     
\begin{center}
\subfigure[$\Lambda_O$ exceptional]{\includegraphics*[angle=0,width=0.6\textwidth]{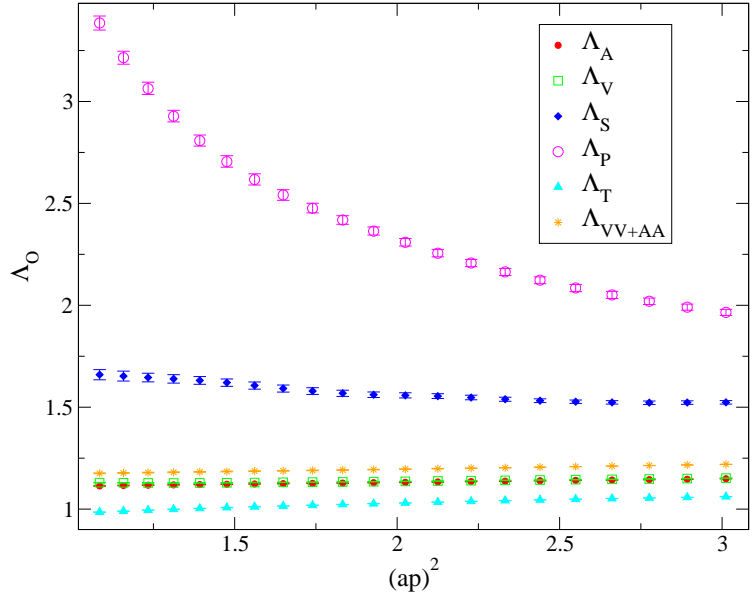}} 
\subfigure[$\Lambda_O$ non-exceptional $\gamma_\mu$ scheme]{
\includegraphics*[angle=0,width=0.6\textwidth]{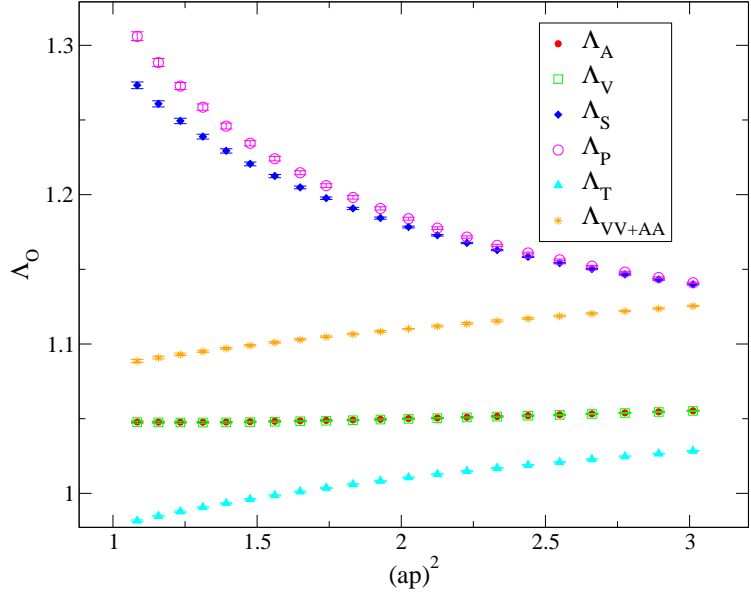}} 
\end{center}
\caption{
All the vertex functions that have been calculated on 
the $\beta = 2.13$ lattice. Non-exceptional and exceptional vertex functions in the chiral limit.
Note the great reduction in chiral symmetry breaking effects 
($\Lambda_A - \Lambda_V$ and $\Lambda_S - \Lambda_P$) at non-exceptional momentum.
\label{fig:cfall}                 
}
\end{figure}

Figure~\ref{fig:cfall} displays the vertex functions of all the operators
analysed in
this paper (bilinear Vector, Axial Vector, Scalar, Pseudoscalar, Tensor and 
four quark operator ${\cal O}_{VV+AA}$) for the $\beta=2.13$ ensemble, for both non-exceptional and exceptional momenta. We will return study each of these structures in turn and in more detail below.

\begin{figure}[htb]     
\begin{center}
\subfigure[$\Lambda_A$, $\Lambda_V$ exceptional]{\includegraphics*[angle=0,width=0.6\textwidth]{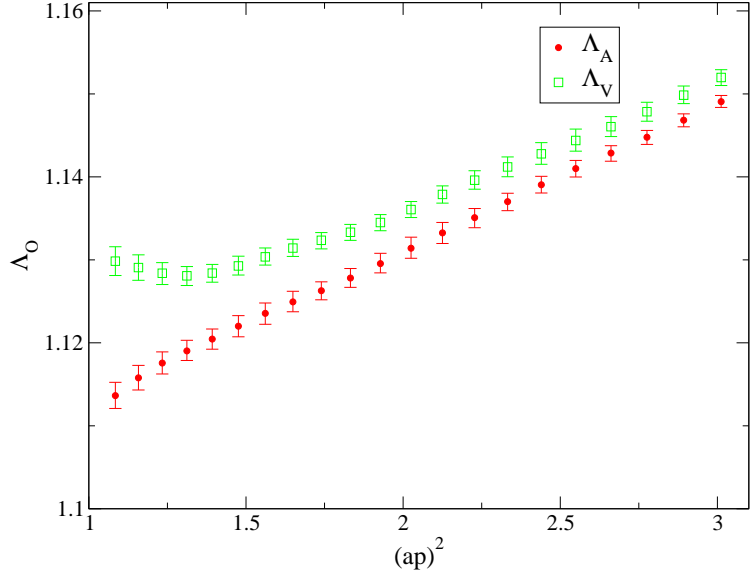}} 
\subfigure[$\Lambda_A$, $\Lambda_V$ non-exceptional $\gamma_\mu$ scheme]{
\includegraphics*[angle=0,width=0.6\textwidth]{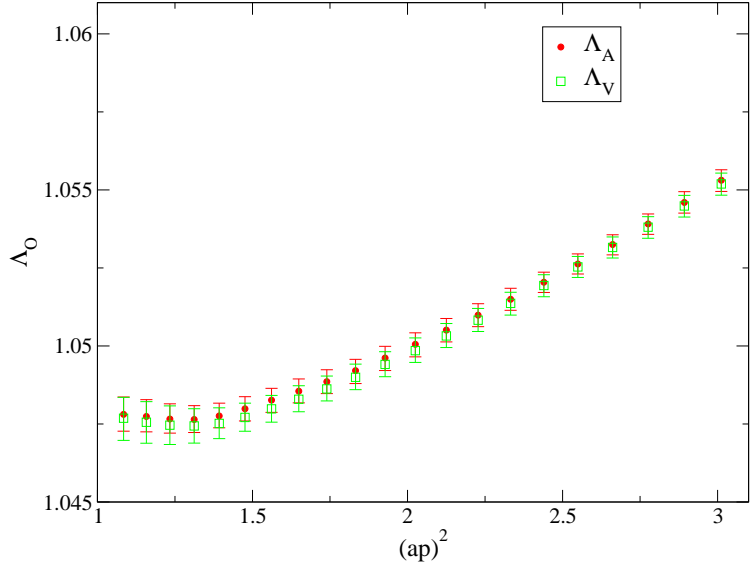}} 
\end{center}
\caption{
Zoomed in view of $\Lambda_A$ and $\Lambda_V$ in the chiral limit with 
non-exceptional and exceptional kinematics from figure \ref{fig:cfall}. 
Note the expanded scale in the non-exceptional plot. 
The roughly 2\% difference in the exceptional case is eliminated, 
even at the lowest momenta, for non-exceptional kinematics.
\label{fig:cfVA}                 
}
\end{figure}

We wish to eliminate the wavefunction renormalisation by taking ratios
$\frac{\Lambda_A}{\Lambda_O}$ and where $\Lambda_A$ is the vertex function
of the local axial current, and its renormalisation is eliminated using
$Z_A$ previously determined from the ratio of matrix elements the local
axial current and the conserved axial current 
From the Ward identities in the limit of small mass and large momentum we should find that 
$ Z_V \simeq Z_A $ and $Z_S = Z_P$. These are well supported by our data for non-exceptional
momentum but, over our momentum range, not held for exceptional momentum.
In order to compare with previous results we adopt,  for exceptional momentum, the
the strategy of \cite{Aoki:2007xm}
eliminating the quark field renormalisation using
\begin{equation}
 \frac{Z_q}{Z_A} = \frac{1}{2} (\Lambda_A + \Lambda_V),
\end{equation}
where factors of $Z_A$ are then multiplied out using a previously calculated value. 
For exceptional data this matches central value conventions 
with \cite{Aoki:2007xm}.
The 2\% difference between $\frac{1}{2} (\Lambda_A + \Lambda_V)$ and $\Lambda_A$ is a chiral 
symmetry breaking uncertainty that propagates globally in operator renormalisation analysis as
a systematic uncertainty (and is doubled for four quark operators).
This uncertainty is eliminated for the non-exceptional momentum kinematic \ref{fig:cfVA}.

For the non-exceptional case  we use simply $\frac{Z_q}{Z_A} = \Lambda_A$.
In this case, we have
\begin{eqnarray}
{Z_m Z_A} &=& \frac{\Lambda_S }{\Lambda_A}, \\
\frac{Z_T}{Z_A} &=& \frac{\Lambda_A }{\Lambda_T}, \\
Z_{B_K} = \frac{Z_{{\cal O}_{VV+AA}}}{Z_A^2} &=& \frac{\Lambda_A^2 }{\Lambda_{{\cal O}_{VV+AA}}}.
\end{eqnarray}

\begin{figure}[htp]
\begin{center}
\subfigure[$Z_q^{RI/MOM}((ap)^2)$ $\rightarrow$ $Z_q^{\overline{MS}}(2 {\rm GeV})$ ]{\label{fig:Zq-a}
\includegraphics*[angle=0,width=0.45\textwidth]{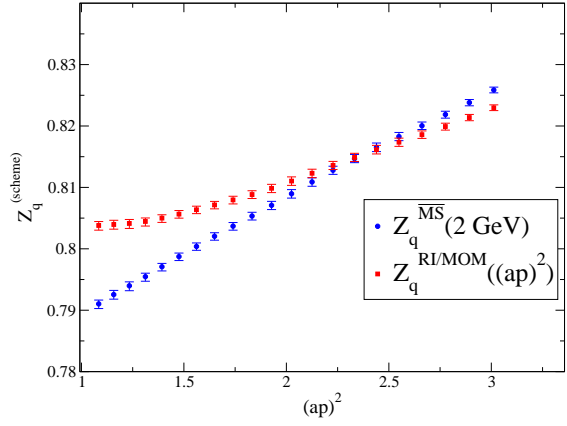}} \\
\subfigure[$Z_q^{SMOM-\cancel{q}}((ap)^2)$ $\rightarrow$ $Z_q^{\overline{MS}}(2 {\rm GeV})$ ]{\label{fig:Zq-b}
\includegraphics*[angle=0,width=0.45\textwidth]{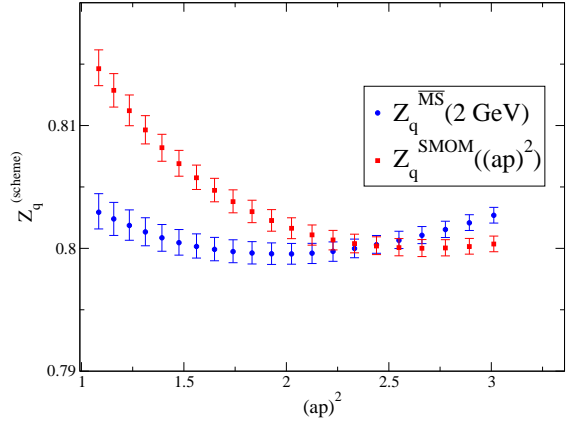}} \\
\subfigure[$Z_q^{SMOM-\gamma_\mu}((ap)^2)$ $\rightarrow$ $Z_q^{\overline{MS}}(2 {\rm GeV})$ ]{\label{fig:Zq-c}
\includegraphics*[angle=0,width=0.45\textwidth]{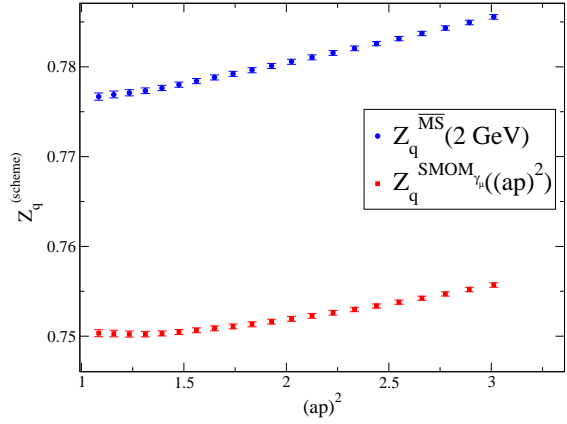}}
\end{center}
  \caption{
\label{fig:2.13Zq}
$Z_q$ on the $\beta = 2.13$ lattice in a MOM scheme (red) and with the 
perturbative running divided out and converted to $\overline{MS}$ (blue).
$Z_q$ contains strong lattice artefacts
for DWF and is opposite to the continuum running.
We find later that this becomes consistent in the continuum limit.
}
\end{figure}

Our results for  RIMOM, SMOM-$\cancel{q}$, and SMOM-$\gamma_\mu$ schemes 
for the wave function renormalisation determined from the axial current vertex
function (non-exceptional) and average of axial and vector vertex functions (exceptional)
are displayed in figure~\ref{fig:2.13Zq}. Here, the running at this lattice spacing
is poorly described by continuum perturbation theory, and is associated with the momentum
dependence of the exponent for binding of light modes in the fifth dimension in the domain
wall formulation \cite{Shamir:1993zy}. We note that this $Z_q$
is cancelled in the ratios above when treating other operators.
Naturally, one expects that such discretisation effects will be removed if a controlled continuum
limit is taken, and this will be revisited in later sections.

\begin{figure}[htp]
\begin{center}
\subfigure[$Z_m^{RI/MOM}((ap)^2)$ $\rightarrow$ $Z_m^{\overline{MS}}(2 {\rm GeV})$]{\label{fig:Zm-a}
\includegraphics*[angle=0,width=0.48\textwidth]{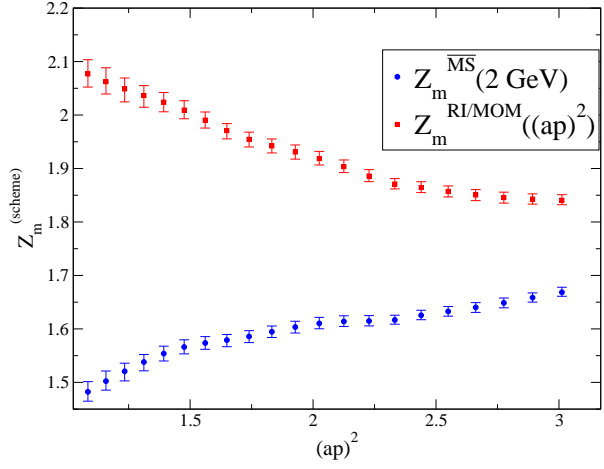}} \\
\subfigure[$Z_m^{SMOM-\cancel{q}}((ap)^2)$ $\rightarrow$ $Z_m^{\overline{MS}}(2 {\rm GeV})$ ]{\label{fig:Zm-b}
\includegraphics*[angle=0,width=0.48\textwidth]{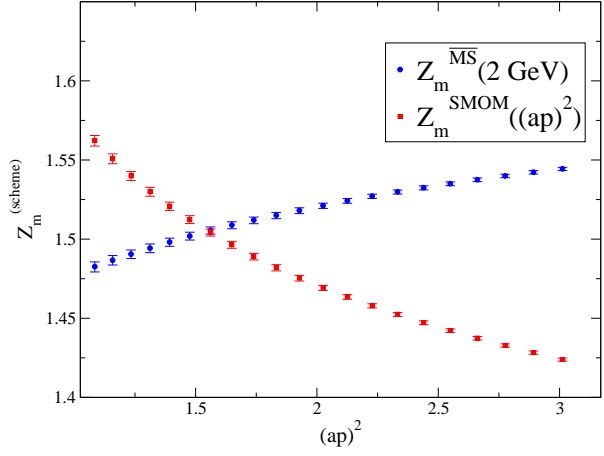}} \\
\subfigure[$Z_m^{SMOM-\gamma_\mu}((ap)^2)$ $\rightarrow$ $Z_m^{\overline{MS}}(2 {\rm GeV})$ ]{\label{fig:Zm-c}
\includegraphics*[angle=0,width=0.48\textwidth]{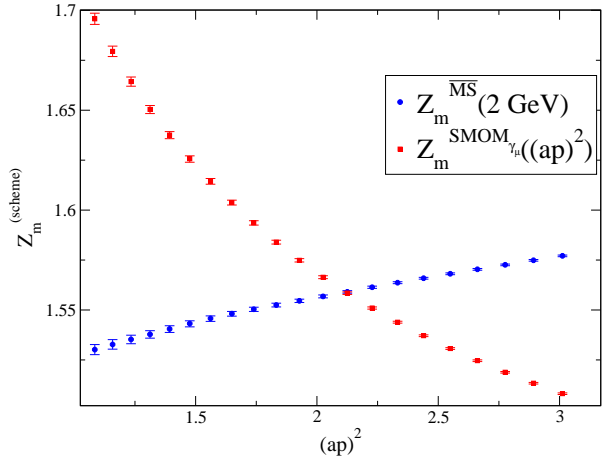}} 
  \end{center}
  \caption{
$Z_m$ on the $\beta = 2.13$ lattice in a MOM scheme (red) and with the 
perturbative running divided out and converted to $\overline{MS}$ (blue).
  \label{fig:2.13Zm}
}
\end{figure}

\begin{figure}[htp]
\begin{center}
\subfigure[$Z_T^{RI/MOM}((ap)^2)$ $\rightarrow$ $Z_T^{\overline{MS}}(2 {\rm GeV})$]{\label{fig:ZT-a}
\includegraphics*[angle=0,width=.48\textwidth]{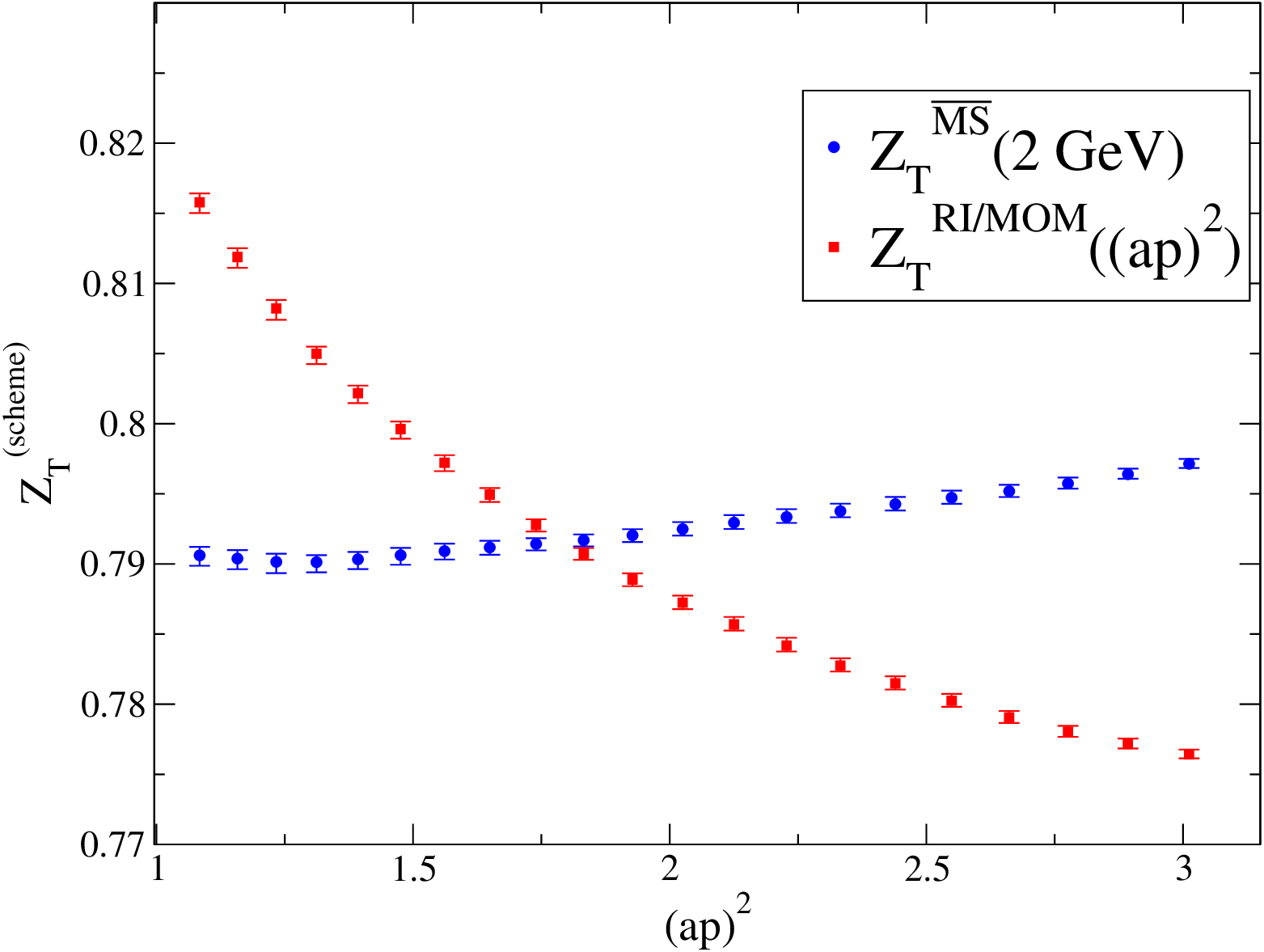}} \\
\subfigure[$Z_T^{SMOM-\cancel{q}}((ap)^2)$ $\rightarrow$ $Z_T^{\overline{MS}}(2 {\rm GeV})$ ]{\label{fig:ZT-b}
\includegraphics*[angle=0,width=.48\textwidth]{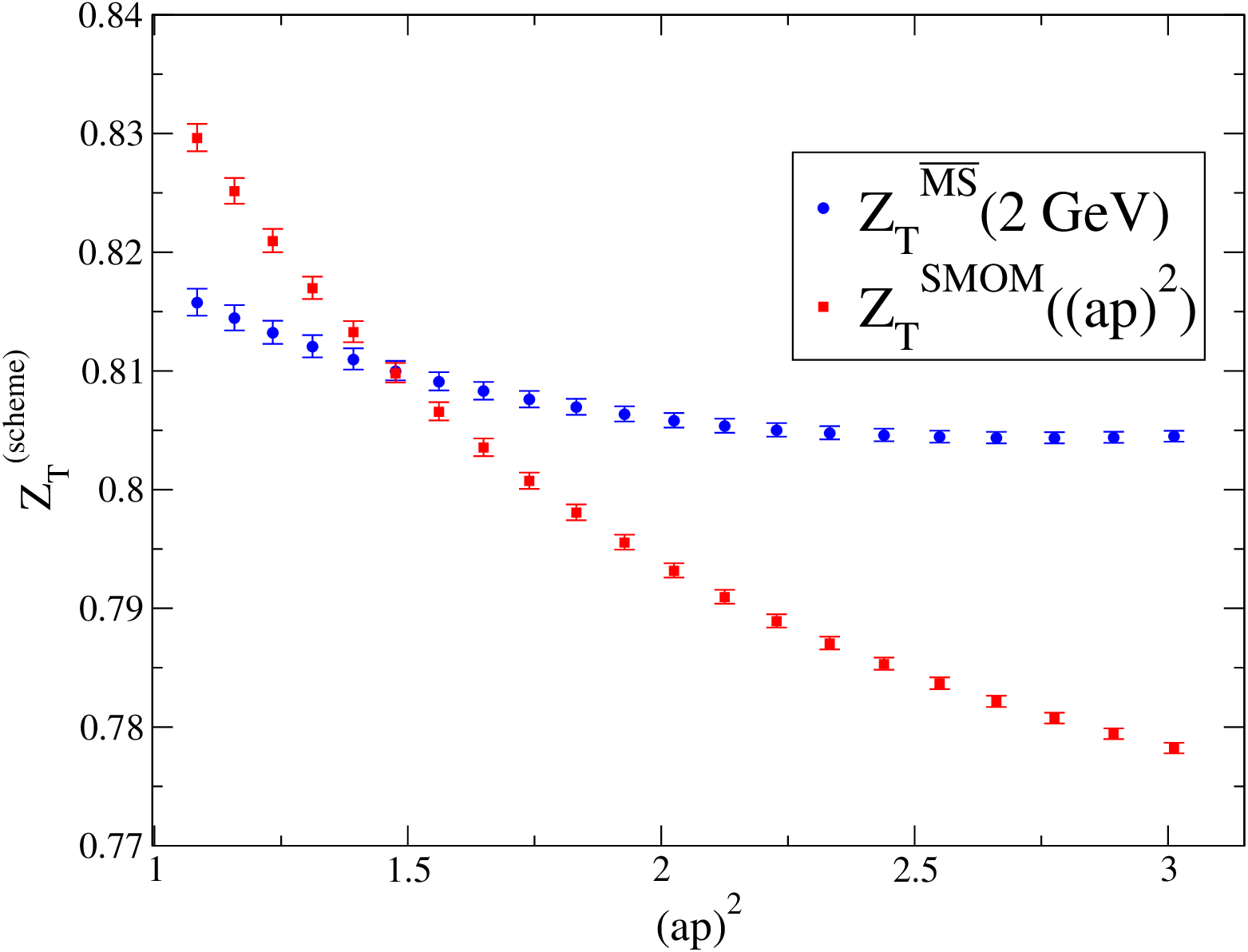}} \\
\subfigure[$Z_T^{SMOM-\gamma_\mu}((ap)^2)$ $\rightarrow$ $Z_T^{\overline{MS}}(2 {\rm GeV})$ ]{\label{fig:ZT-c}
\includegraphics*[angle=0,width=.48\textwidth]{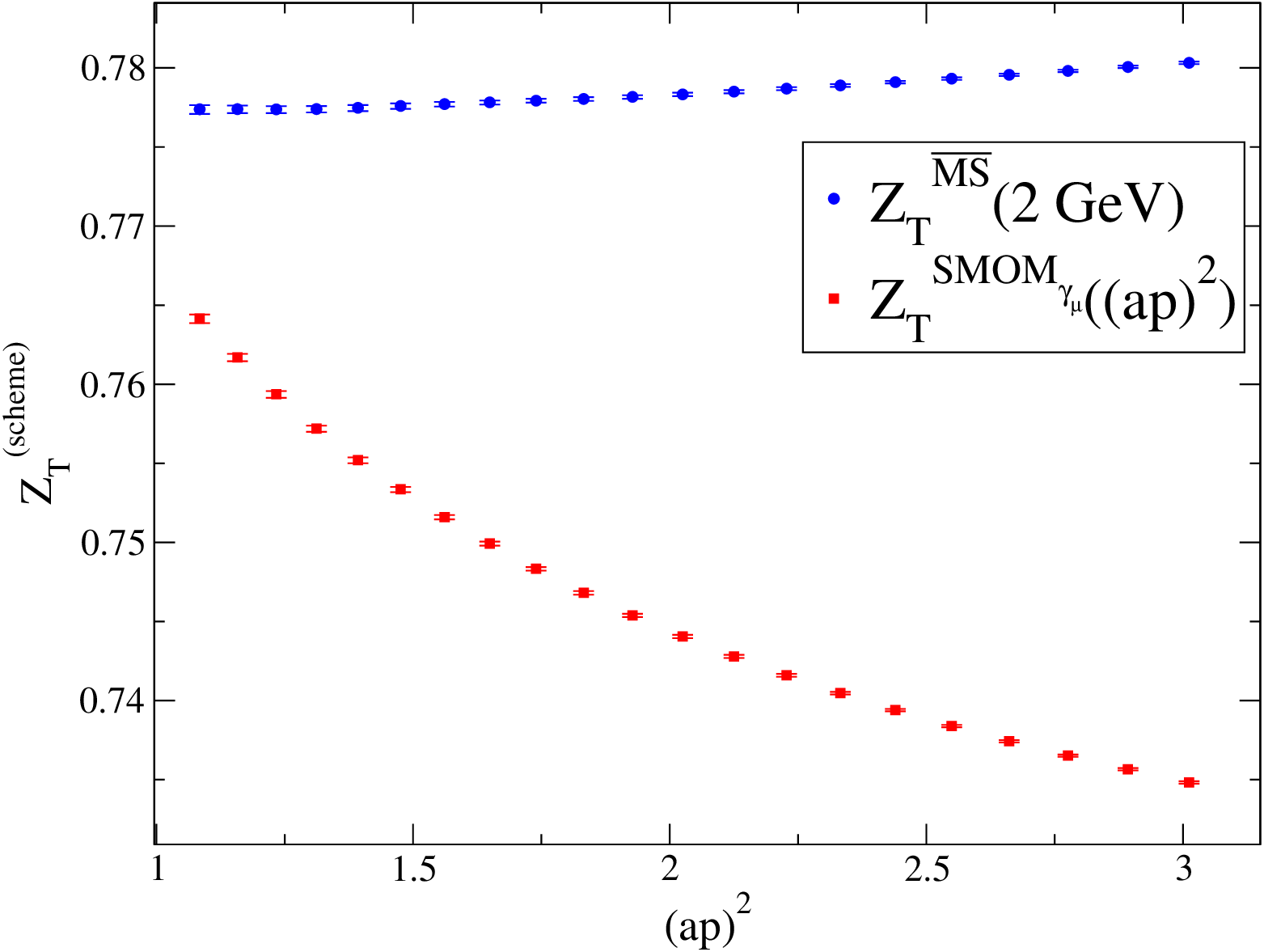}} 
\end{center}
\caption{
$Z_T$ on the $\beta = 2.13$ lattice in a MOM scheme (red) and with the 
perturbative running divided out and converted to $\overline{MS}$ (blue).
  \label{fig:2.13ZT}}
\end{figure}

\begin{figure}[htp]
\psfrag{RI/MOM}{\LARGE  RI/MOM}
\psfrag{SMOMqq}{\LARGE  SMOM$(\slashed{q},\slashed{q})$}
\psfrag{SMOMqg}{\LARGE  SMOM$(\slashed{q},\gamma_\mu)$}
\psfrag{SMOMgg}{\LARGE  SMOM$(\gamma_\mu,\gamma_\mu)$}
\psfrag{SMOMgq}{\LARGE  SMOM$(\gamma_\mu,\slashed{q})$}

\begin{center}
\subfigure[$Z_{BK}^{RI/MOM}((ap)^2)$ $\rightarrow$ $Z_{BK}^{\overline{MS}}(2 {\rm GeV})$]{\label{fig:Zvvpaa-a}
\includegraphics*[angle=0,width=0.48\textwidth]{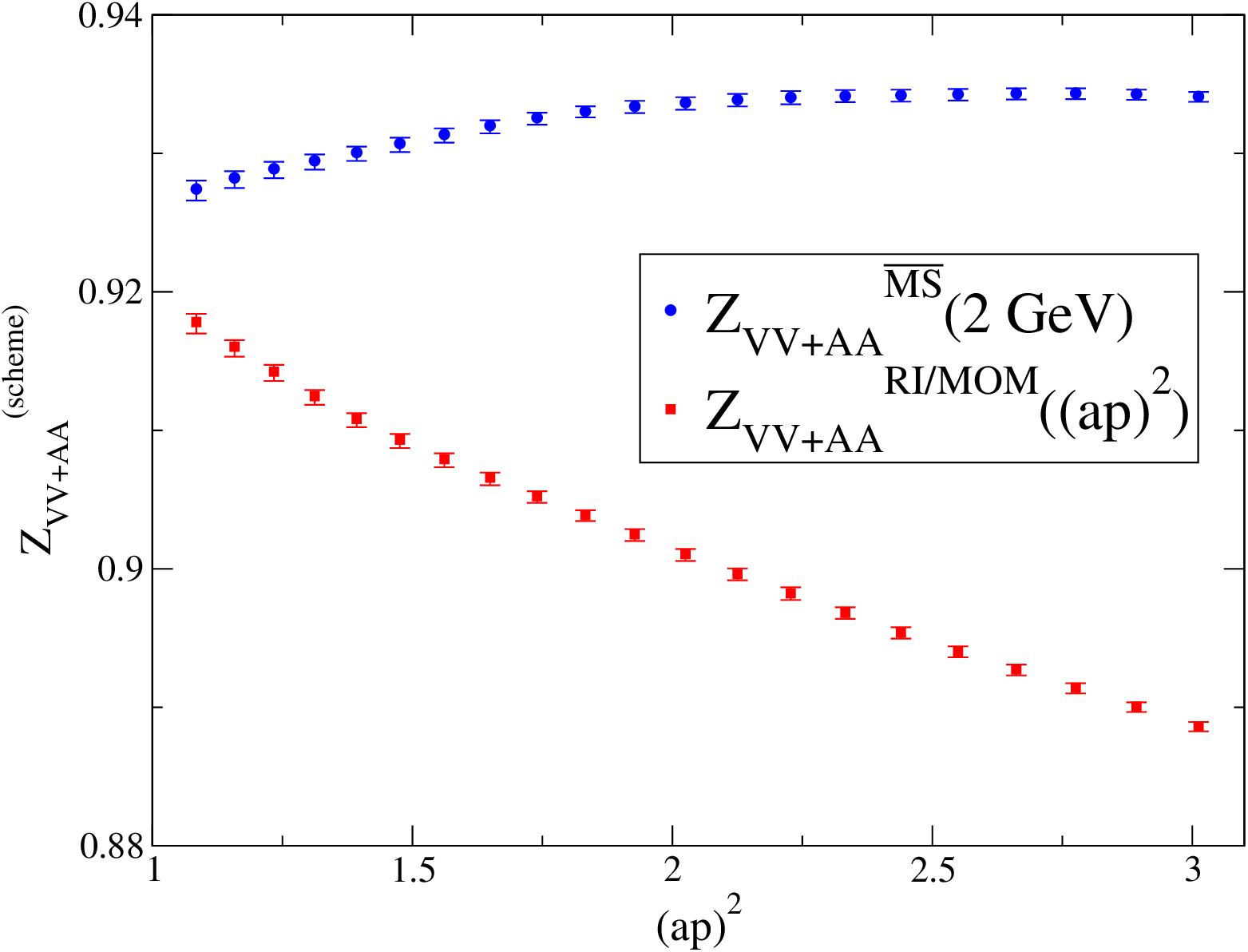}} 
\subfigure[$Z_{BK}^{SMOM-\cancel{q}-\cancel{q}}((ap)^2)$ $\rightarrow$ $Z_{BK}^{\overline{MS}}(2 {\rm GeV})$ ]{\label{fig:Zvvpaa-b}
\includegraphics*[angle=0,width=0.48\textwidth]{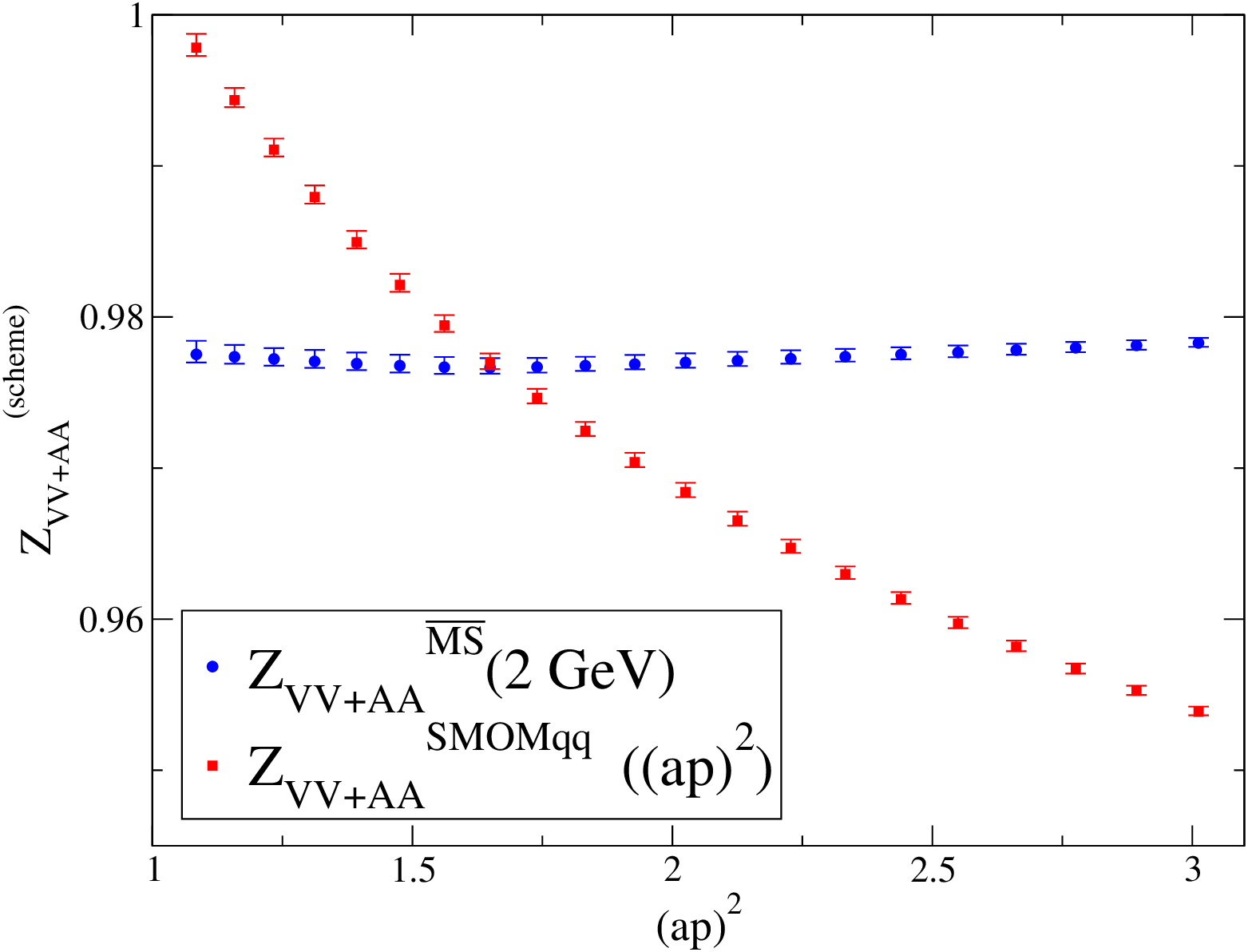}}
\subfigure[$Z_{BK}^{SMOM-\gamma_\mu-\gamma_\mu}((ap)^2)$ $\rightarrow$ $Z_{BK}^{\overline{MS}}(2 {\rm GeV})$ ]{\label{fig:Zvvpaa-c}
\includegraphics*[angle=0,width=0.48\textwidth]{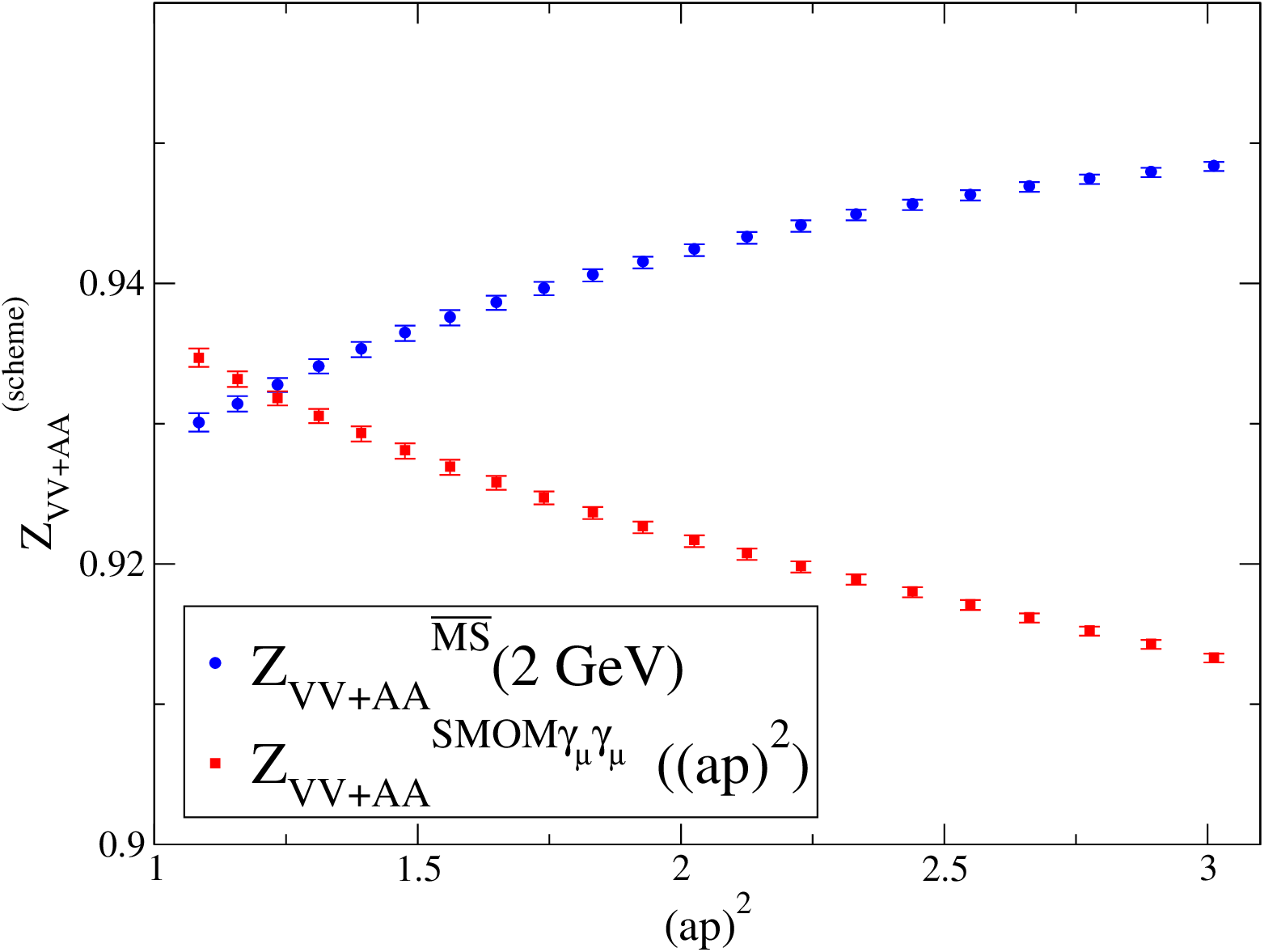}}
\subfigure[$Z_{BK}^{SMOM-\cancel{p}-\gamma_\mu}((ap)^2)$ $\rightarrow$ $Z_{BK}^{\overline{MS}}(2 {\rm GeV})$ ]{\label{fig:Zvvpaa-d}
\includegraphics*[angle=0,width=0.48\textwidth]{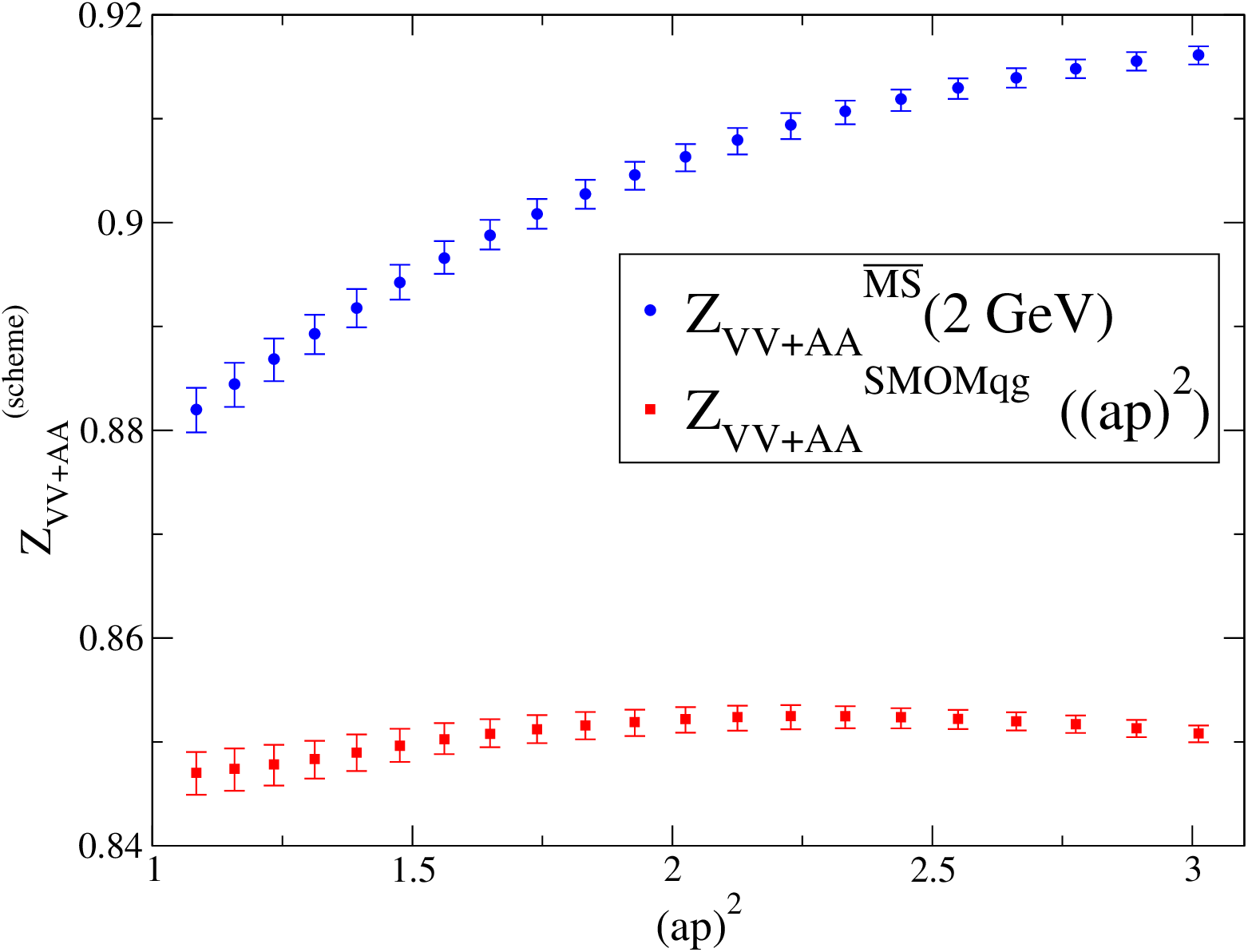}}
\subfigure[$Z_{BK}^{SMOM-\gamma_\mu-\cancel{p}}((ap)^2)$ $\rightarrow$ $Z_{BK}^{\overline{MS}}(2 {\rm GeV})$ ]{\label{fig:Zvvpaa-e}
\includegraphics*[angle=0,width=0.48\textwidth]{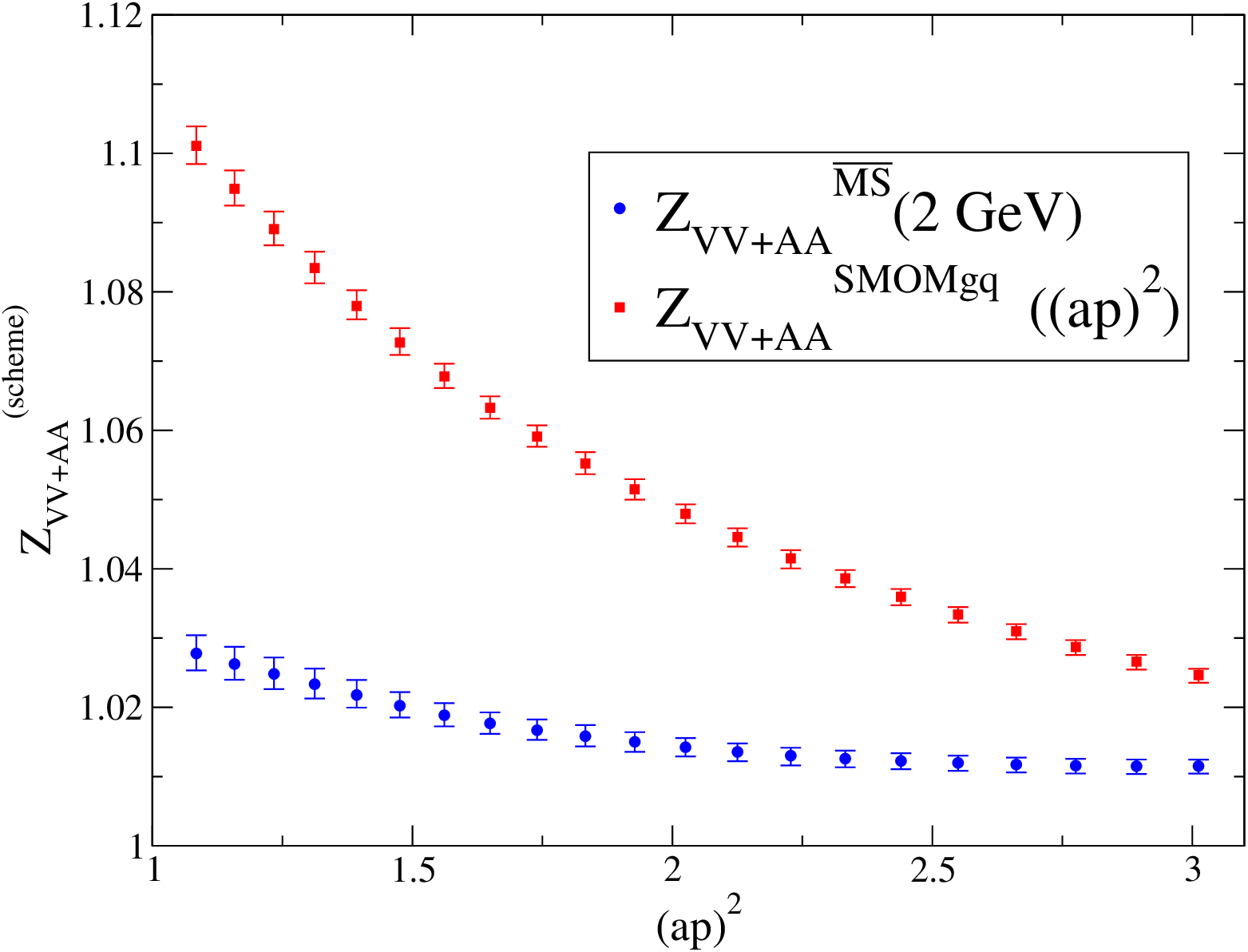}}
\end{center}
\caption{
$Z_{VV+AA}$ on the $\beta = 2.13$ lattice in a MOM scheme (red) and with the 
perturbative running divided out and converted to $\overline{MS}$ (blue).
\label{fig:2.13Zvvpaa}
}
\end{figure}

Figures~\ref{fig:2.13Zm}, \ref{fig:2.13ZT}, and \ref{fig:2.13Zvvpaa} display the corresponding
data for the mass, tensor current and four quark operator ${\cal O}_{VV+AA}$ relevant for
$B_K$. As promised, the data is extremely precise and ${\cal O}(4)$ breaking discretisation
effects do not introduce scatter in the data as the momentum is changed because we are
always selecting the same lattice momentum with our twisted boundary conditions. 
Errors are at the 0.1\% level even on this relatively small volume and
with only twenty configurations.

In contrast to $Z_q$, the running of $Z_m$, $Z_T$, and $Z_{VV+AA}$ 
is reasonably well described 
by continuum perturbation theory for larger values of $p^2$, even on this non-zero lattice spacing. 
This is indicated by reduction of the slope of the data
after perturbative conversion to $\overline{MS}$ (blue) especially at higher
momenta. Residual $p^2$ dependence remains after this conversion
at the few percent scale. This reflects an admixture of lattice artefacts, and
perturbative truncation error. The methods introduced in this paper enable
the continuum limit of the scale evolution to be determined and disentangle these
sources of error.

Table \ref{tab:result_cf} shows our results compared with the results of Ref \cite{Aoki:2007xm} after
following the same procedure of converting to
 the $\overline{MS}$ scheme at 2 GeV and then 
extrapolating to $p^2=0$. In all cases the extrapolation was
performed using a linear fit of $\Lambda_O$ as a function 
of $(ap)^2$ in the range $2 < (ap)^2 < 3.2$. 
This procedure is flawed. We note that this is in danger of 
extrapolating perturbative errors into the infrared. 
In the next section we show instead how to take the continuum limit at non-zero
momentum to eliminate discretisation errors. 
These numbers are thus for comparison only, and constitute 
a check and a demonstration of the precision 
that is possible using the techniques of this paper.
The numbers do not represent an attempt to improve on previous estimates of 
renormalization constants for the RBC-UKQCD domain wall programme. 
The physical RBC-UKQCD predictions will be updated using the methods of
this paper techniques in later works, where we
will also use step scaling to raise the momentum scale to 
minimise perturbative error.

\begin{table}[hbt]
\begin{tabular}{cccccc}
\hline
\hline
$Z_O$ & Ref: \cite{Aoki:2007xm} & RIMOM (extrap) & RIMOM & SMOM-$\cancel{q}$ & SMOM-$\gamma_\mu$ \\
\hline
$Z_q$   & 0.7726(48 + 150) & 0.7753(15) & 0.79605(63) & 0.8035(10) & 0.77744(31)\\
$Z_m$   & 1.656(30 + 83) & 1.483(28) & 1.546(15) & 1.5073(26) & 1.5405(18)\\
$Z_T$   & 0.7950(34 + 150)  & 0.7962(13) & 0.80339(62) & 0.80626(73) & 0.77768(20)\\
\hline
\end{tabular}
\caption{\label{tab:result_cf}
The quark field, mass and tensor current renormalization constants in $\overline{MS}$ at $2(GeV)$. Error in the first column is (stat + sys) error in the other columns statistical only. Note this work used 20 configurations at each mass whereas \cite{Aoki:2007xm} used 300 point source measurements on 75 configurations. The third column uses the value extrapolated to zero for comparison with the results of \cite{Aoki:2007xm}. The others use simple interpolation to obtain the value at $p^2 = 2 {GeV}$ }
\end{table}

\begin{center}
\begin{table}[hbt]
\begin{tabular}{ccccccc}
\hline
\hline
Ref: \cite{Aoki:2007xm} & RIMOM (extrap) & RIMOM & SMOM-($\cancel{q}$,$\cancel{q}$) & SMOM-($\gamma_\mu$,$\gamma_\mu$) &  SMOM-($\cancel{q}$,$\gamma_\mu$) & SMOM-($\gamma_\mu$,$\cancel{q}$) \\
\hline
0.9276(52 + 220) & 0.93330(73) & 0.92994(54) & 0.97737(59) & 0.93406(60) & 1.0233(21) & 0.8903(19) \\
\hline
\end{tabular}
\caption{\label{tab:result_cfbk}
BK renormalization constant with same parameters as above. The extrapolated RIMOM is for comparison with \cite{Aoki:2007xm} the rest of the measurements use the interpolated value at $p^2 = 2 {GeV}$. }
\end{table}
\end{center}

\subsection{Scale determination in a small volume}

Determining the lattice spacing for our $\beta=2.23$ ensemble is a useful
test case for the methods of section~\ref{sec:traj}.
As a preliminary we have calculated $r_C$ for our $16^3$ lattices. We compute 
timelike Wilson loops with four hits of APE smearing, smearing parameter 2.5, 
in the spatial direction. The tree level improvement of \cite{Necco:2001xg} is 
here required for the Iwasaki gluon action. We compute the tree level Wilson 
loop using code developed in \cite{Bali:2002wf} to obtain Table \ref{tab:Vglue} where the potential 
includes the self energy part.
\begin{equation}
aV(\vec{R}a) = - \lim_{T\to \infty} \frac{1}{W(\vec{R}, T)} \frac{d W(\vec{R}, T)}{dT} = C_FV(\vec{R}) g^2 + O(g^4)
\end{equation}

\begin{table}[hbt]
\begin{tabular}{ccc}
$\vec{R}$ & $V_{W}$ & $V_{I}$ \\
\hline
(1,0,0) & 0.166667          & 0.08963(1)\\
(2,0,0) & 0.209842          & 0.12569(2)\\
(3,0,0) & 0.225186          & 0.13879(3)\\
(4,0,0) & 0.232442          & 0.14506(4)\\
(5,0,0) & 0.236630          & 0.14878(4)\\
(6,0,0) & 0.239366          & 0.15133(5)\\
(7,0,0) & 0.241300          & 0.15318(5)\\
(8,0,0) & 0.242742          & 0.15456(5)\\
\hline
\end{tabular}
\caption{\label{tab:Vglue}
$V_W$ is the static potential tree level part with the Wilson gluon action, 
$V_I$ uses the Iwasaki gluon action. As $R$ tends to infinity the difference 
between successive terms for Wilson and Iwasaki actions (the force) is the same 
since the self energy part cancels and the two actions reproduce the same IR physics.
}
\end{table}

By demanding
\begin{equation}
F(r_I) = (V(r) - V(r-a))/a = F_{tree} = \frac{1}{4 \pi r_I^2}
\end{equation}
we solve for $r_I$ using Table \ref{tab:Vglue}. This approach leads to much reduced 
lattice $O(a^2)$ errors \cite{Necco:2001xg} \cite{Hasenfratz:2001tw}. Further, using 
the force instead of the potential removes the self energy part and reduces the linear 
term, $\sigma r$, to a constant. Fitting the Cornell potential to find $r_C$ requires $\sigma$. 
However, $\sigma$ is dependant on the long distance behaviour of $V(r)$, and to constrain 
it one has to sample large $r$. When fitting to the force, $r_C$ can be obtained without 
including large distance data in the procedure, and $r_C$ computed this way 
is a finite volume safe observable with which to determine the lattice spacing. 

In order to extract $V(r)$ from the average Wilson loops $\langle W(r,t) \rangle$ 
we first plot the effective potential 
\begin{equation}\label{logeffm}
\log(\frac{\langle W(r,t+1) \rangle}{\langle W(r,t) \rangle})
\end{equation}
as a function of $t$. The largest value of $r$ used in this analysis is $4a$. 
Excited state contamination and statistical noise are problematic, and 
we use the 'black box' method \cite{Fleming:2004hs} to define an improved effective mass that
takes account of the first excited state. This allows us to extend $t_{min}$ to $t = 3$ 
which gives a significant error reduction compared to 
the usual effective mass. On the $\beta = 2.13$ lattice we use $600$ 
configurations for each mass, each configuration rotated to use 4 different time directions.
For $\beta = 2.23$ we use $750$ configurations per mass and again rotate the time direction.

\begin{figure}[h]
\begin{center}
\includegraphics*[width=100mm]{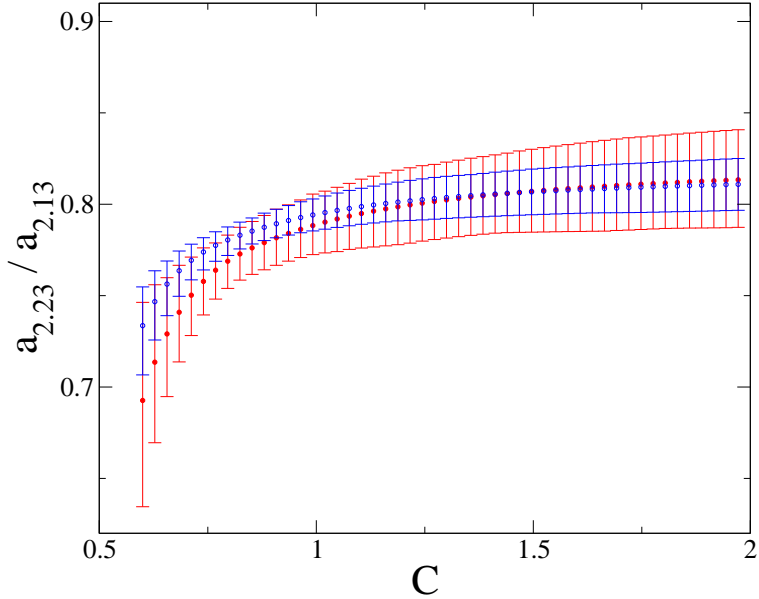}
\end{center}
\caption{The ratio $\frac{r_C / a_{2.13}}{r_C / a_{2.23}}$ as a function of C. 
The red data is computed using the effective mass \ref{logeffm} $t_{min} = 4$ 
and the blue uses the black box method with $t_{min} = 3$, and gives 
the same value but with significantly smaller error.\label{fig:constbb}}
\end{figure}

The equation $r^2 F(r) = C$ has solutions in the range of our data for $C \in (0.6,2.0)$. 
Lower values give more accurate $r_C$ but these are likely to have large 
discretisation errors. 
For any $C$ however $\frac{r_C / a_{2.13}}{r_C / a_{2.23}}$ should be constant, we plot this 
ratio as a function of $C$ in figure \ref{fig:constbb} and determine $C > 1.4$ is 
appropriate and gives the value for the ratio of lattice scales,
\begin{equation}
R_a^2 = \frac{(a_{2.23})^2}{(a_{2.13})^2} = 0.652(21)
\end{equation}
This implies a lattice spacing of $a^{-1}\simeq 2.14$ GeV for the $\beta=2.23$ ensembles.

\begin{figure}[h]
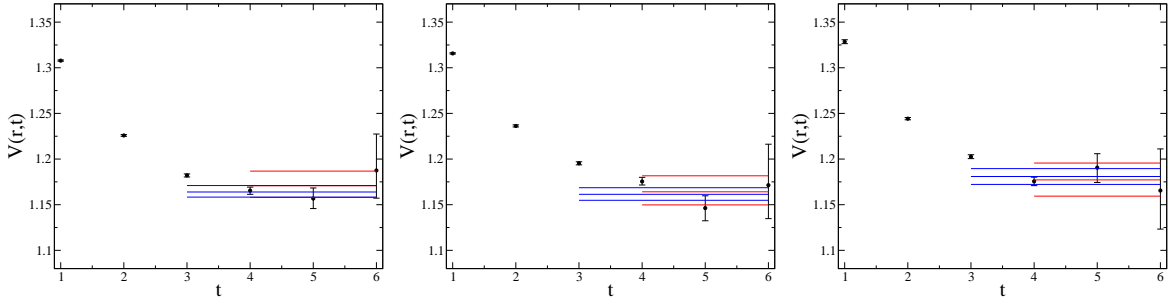

\begin{center}$
\begin{array}{ccc}
\includegraphics*[width=50mm]{effpot0.01.eps} &
\includegraphics*[width=50mm]{effpot0.02.eps} &
\includegraphics*[width=50mm]{effpot0.03.eps}
\end{array}$
\end{center}
\caption{Effective mass plots for (left to right) $m = 0.03$, $m=0.02$, $m = 0.01$ 
on the $\beta=2.13$ lattice at $r = 4a$. 
Datapoints are computed via equation \ref{logeffm}. 
The red constant is obtained from a fit over the last three datapoints. 
The blue is from the black box method with $t_{min} = 3$.}
\end{figure}

We propose below a scale factor $s = 1.5$, and believe that at least
two iterations of step scaling should be possible based only on
the static potential. However, the static potential displays percent
scale errors even with many configurations and finding a more
precise alternative would certainly be good in any case.

\subsection{Continuum limit scale evolution functions}
\label{sec:stepdata} 

Following equation \ref{eq:sig} we can compute $\sigma_O(p,sp)$ using two lattice spacings. 
We consider the evolution of renormalisation constants 
$Z_q$, $Z_m$, $Z_T$, and $Z_{B_K}$  
in both the exceptional and the non-exceptional kinematic schemes.
We choose $p \simeq 2(GeV)$ and compute $\sigma_O(p,sp)$ as a function of s.

\begin{figure}[htp]
  \begin{center}
\subfigure[$Z_q$, exceptional, 
$RI$]{\includegraphics*[angle=0,width=0.47\textwidth]{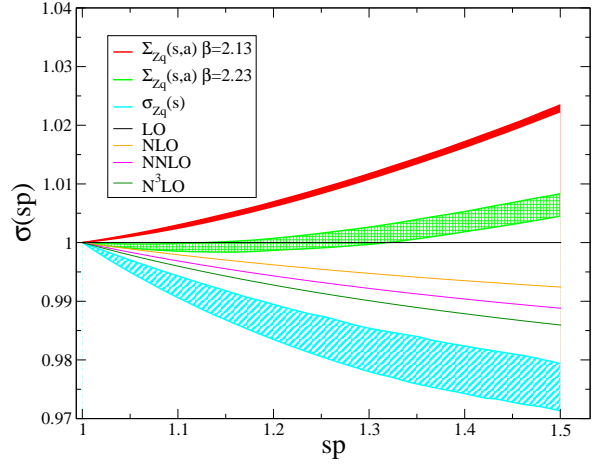}} \\
\subfigure[$Z_q$, non-exceptional, 
$SMOM-\cancel{q}$]{\includegraphics*[angle=0,width=0.478\textwidth]{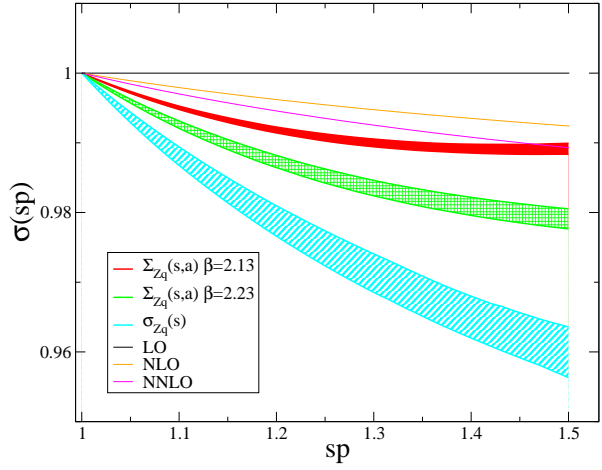}} \\
\subfigure[$Z_q$, non-exceptional, $SMOM-\gamma_\mu$]{\includegraphics*[angle=0,width=0.478\textwidth]{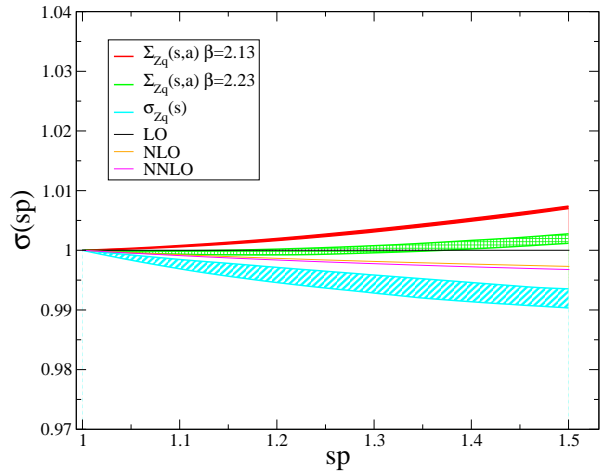}} 

  \end{center}
  \caption{Quark field renormalization running from $\simeq 2$ GeV to $\simeq 3.0$ GeV in the three 
           different schemes compared to the perturbative running in each scheme.
  }
  \label{fig:Zqrun}
\end{figure}

\begin{figure}[htp]
  \begin{center}
    \subfigure[$Z_m$, exceptional, $RI$]{
\includegraphics*[angle=0,width=0.47\textwidth]{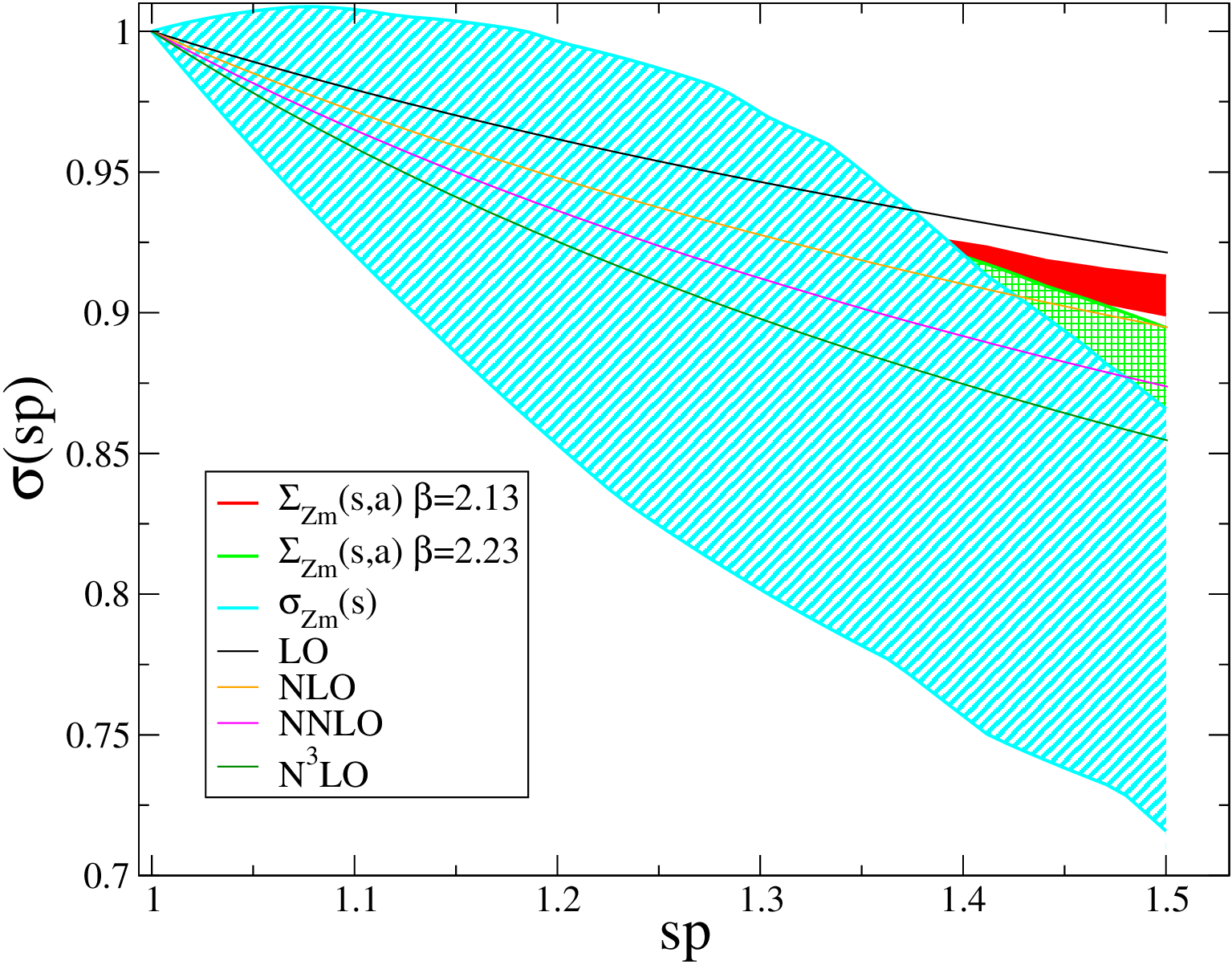}} \\
    \subfigure[$Z_m$, non-exceptional, $SMOM-\cancel{q}$]{\includegraphics*[angle=0,width=0.47\textwidth]{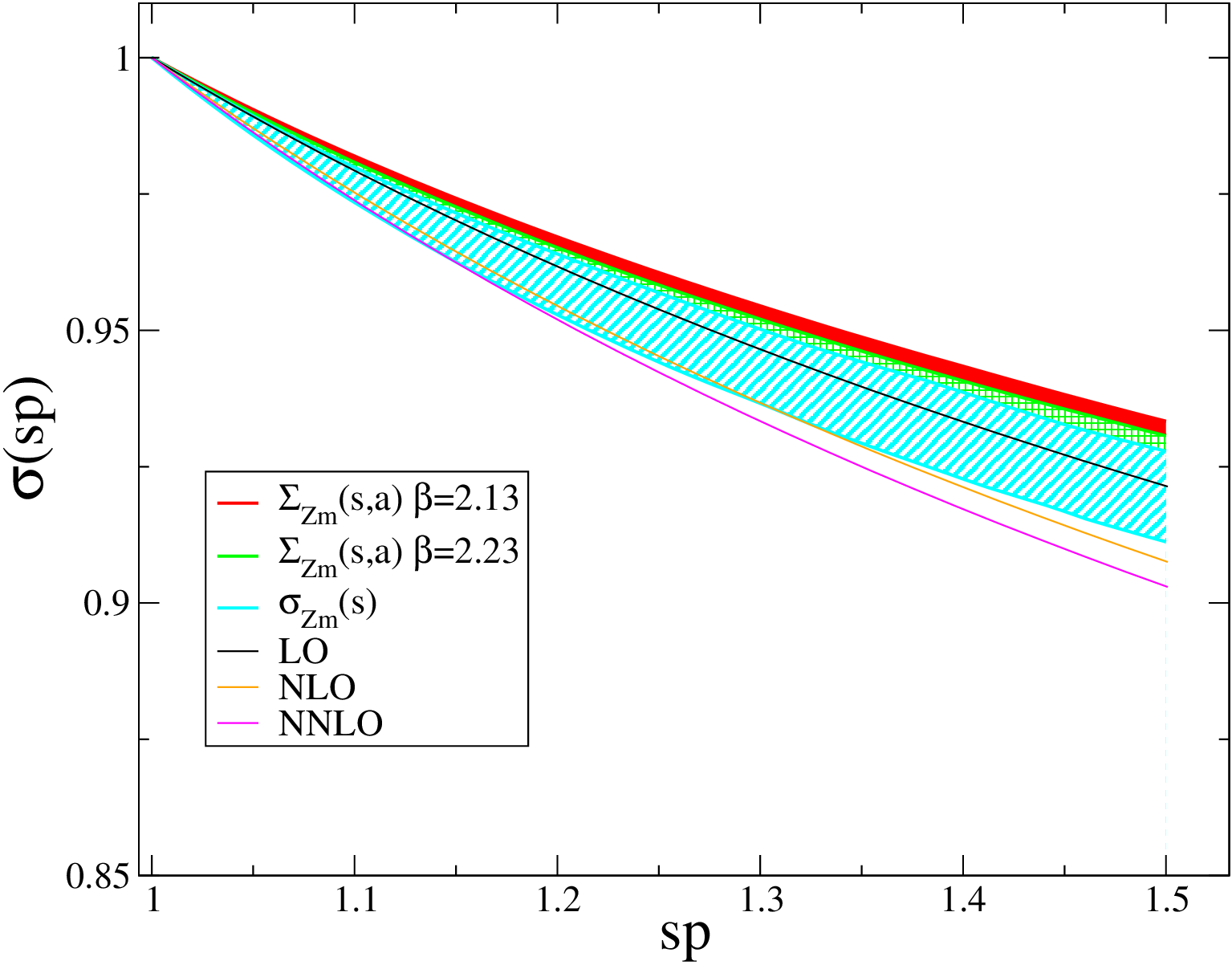}} \\
    \subfigure[$Z_m$, non-exceptional, $SMOM-\gamma_\mu$]{\includegraphics*[angle=0,width=0.47\textwidth]{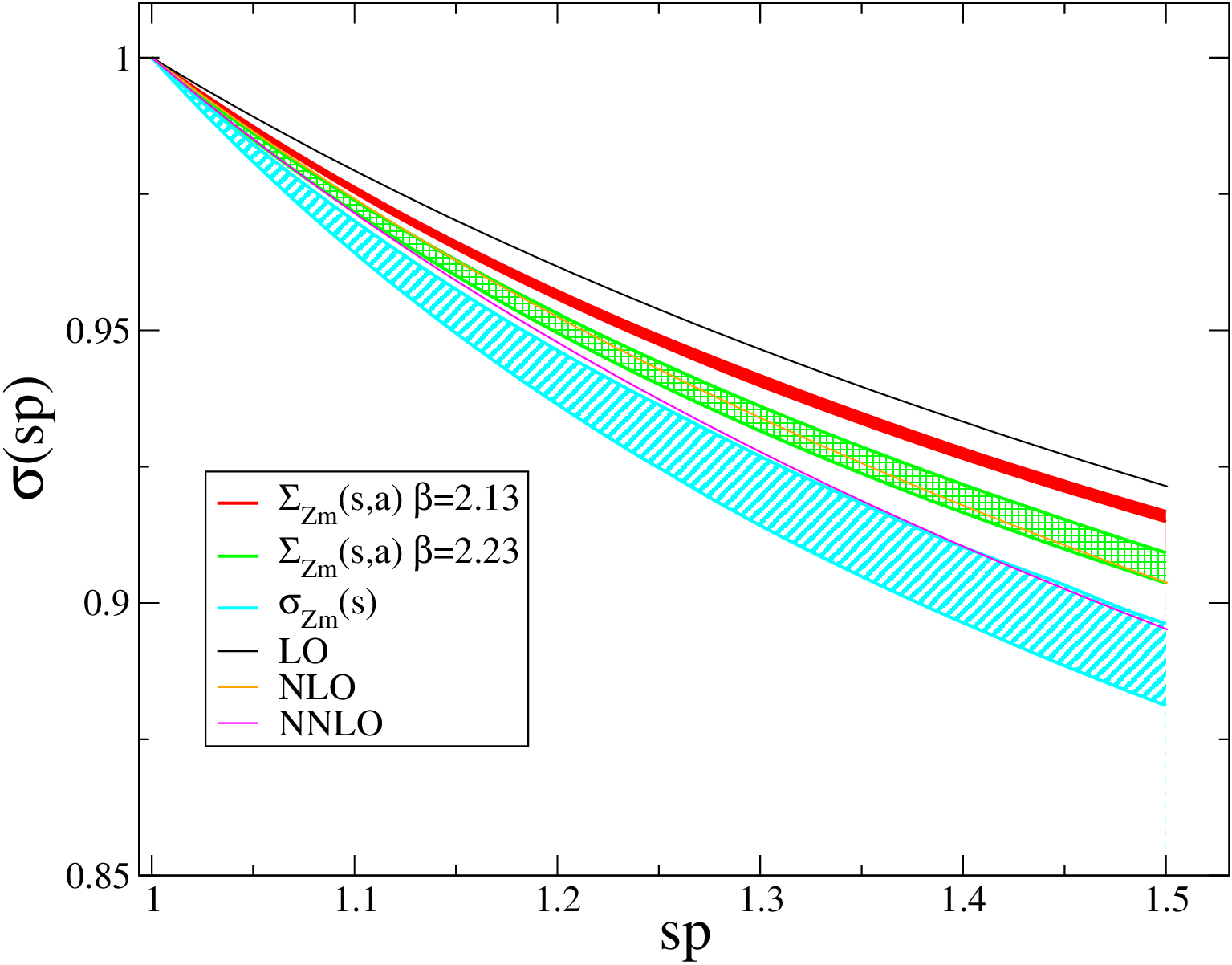}} 

  \end{center}
  \caption{Quark mass renormalization running from $\simeq 2$ GeV to $\simeq 3.0$ GeV in the three different schemes compared to the perturbative running in each scheme.}
  \label{fig:Zmrun}
\end{figure}

\begin{figure}[htp]
  \begin{center}
    \subfigure[$Z_T$, exceptional, $RI$]{\includegraphics*[angle=0,width=0.47\textwidth]{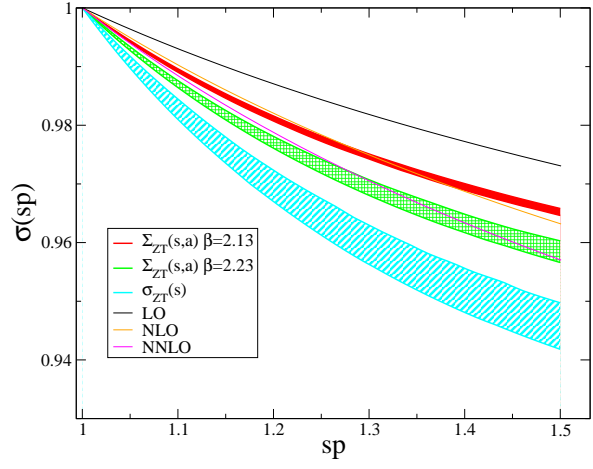}} \\
    \subfigure[$Z_T$, non-exceptional, $SMOM-\cancel{q}$]{\includegraphics*[angle=0,width=0.47\textwidth]{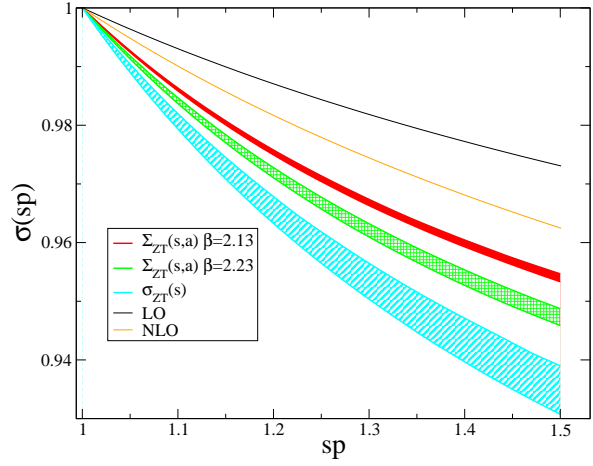}} \\
    \subfigure[$Z_T$, non-exceptional, $SMOM-\gamma_\mu$]{\includegraphics*[angle=0,width=0.47\textwidth]{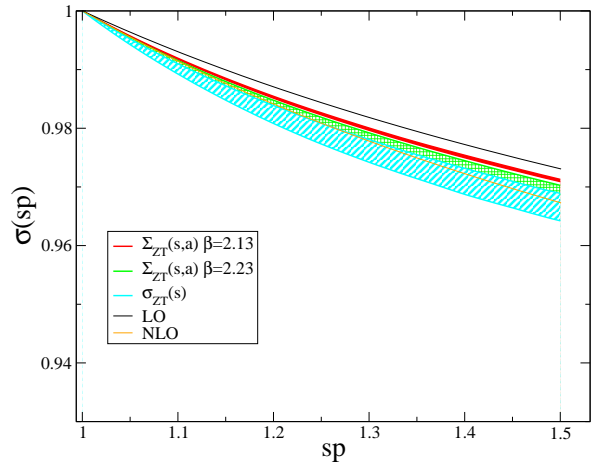}} 

  \end{center}
  \caption{Tensor current renormalization running from $\simeq 2$ GeV to $\simeq 3.0$ GeV in the three different schemes compared to the perturbative running in each scheme.}
  \label{fig:ZTrun}
\end{figure}

Because of the very smooth $p^2$ dependence of the vertex functions computed 
with twisted boundary conditions we can perform a simple interpolation to match 
values of $p^2$ on each lattice. One could in principle simulate at the same physical momentum 
on each lattice by choosing the twisting angles appropriately,
however since the interpolation introduced only a very small uncertainty this was not 
necessary. 
 
\begin{figure}[htp]
  \begin{center}
    \subfigure[$Z_{BK}$, exceptional, $RI/MOM$]{\includegraphics*[angle=0,width=0.45\textwidth]{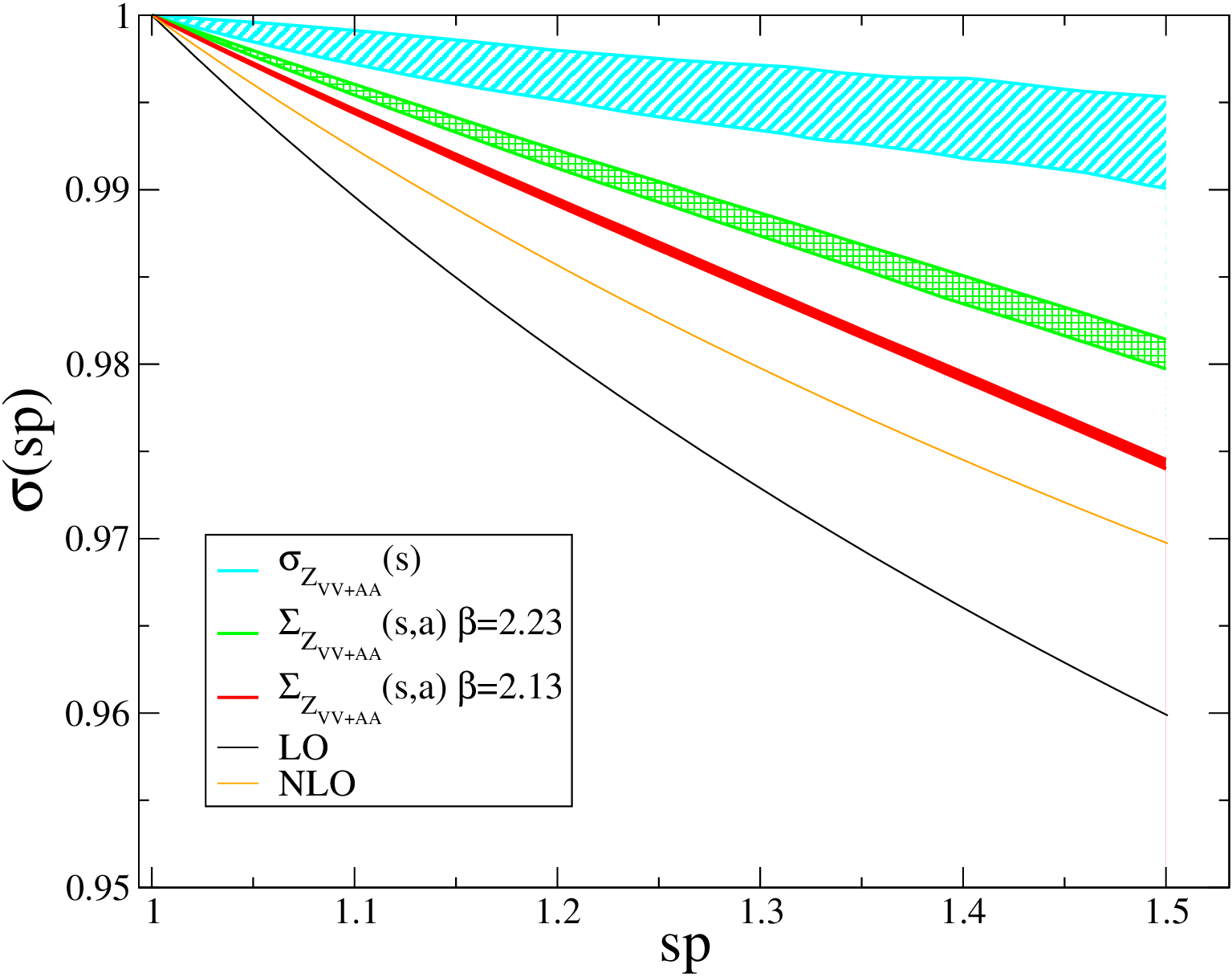}} 
    \subfigure[$Z_{BK}$, non-exceptional, $SMOM-({\cancel{p},\cancel{p}}$)]{\includegraphics*[angle=0,width=0.45\textwidth]{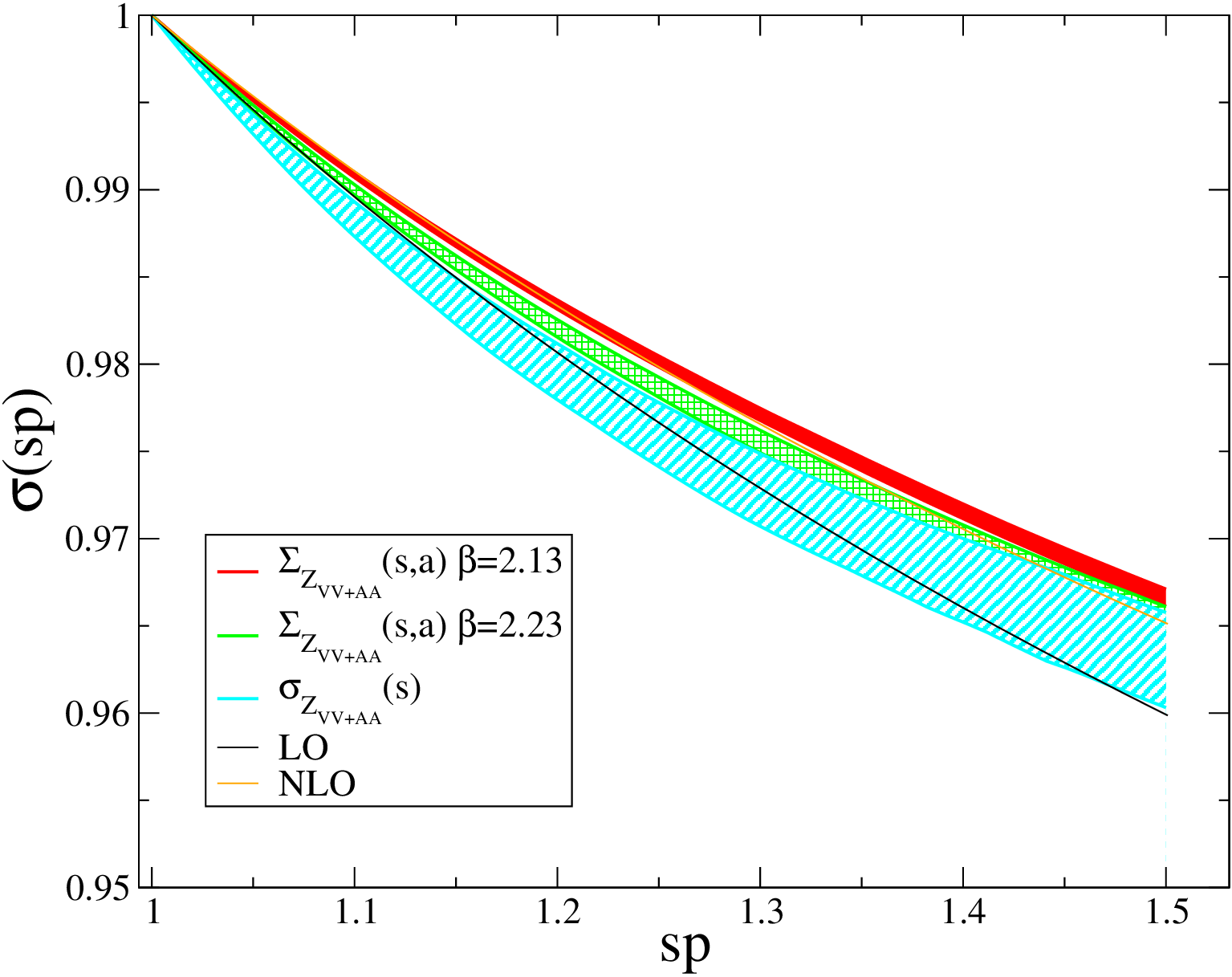}} 
\subfigure[$Z_{BK}$, non-exceptional, $SMOM-({\gamma_\mu,\gamma_\mu}$)]{\includegraphics*[angle=0,width=0.45\textwidth]{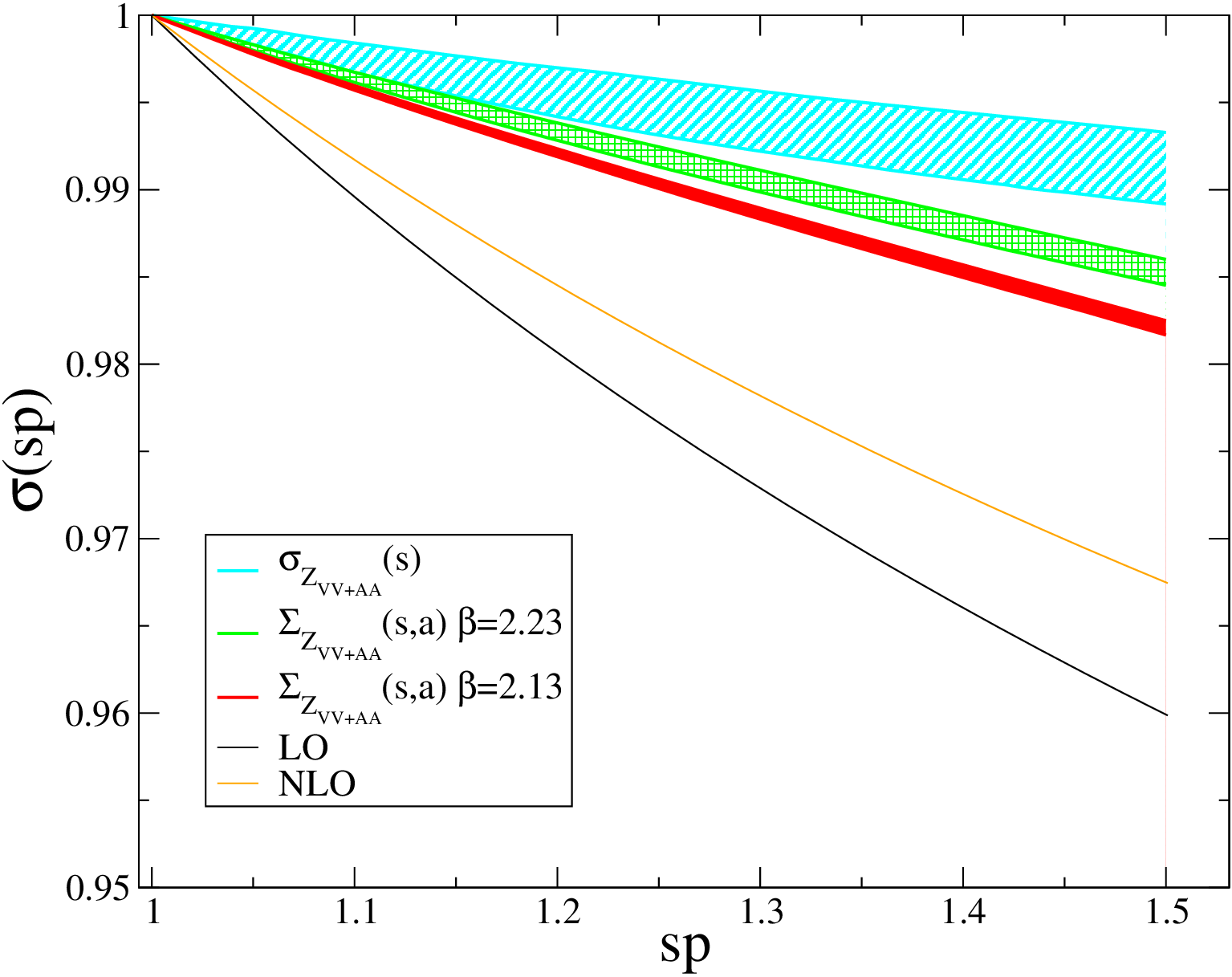}} 
\subfigure[$Z_{BK}$, non-exceptional, $SMOM-({\cancel{p},\gamma_\mu}$)]{\includegraphics*[angle=0,width=0.45\textwidth]{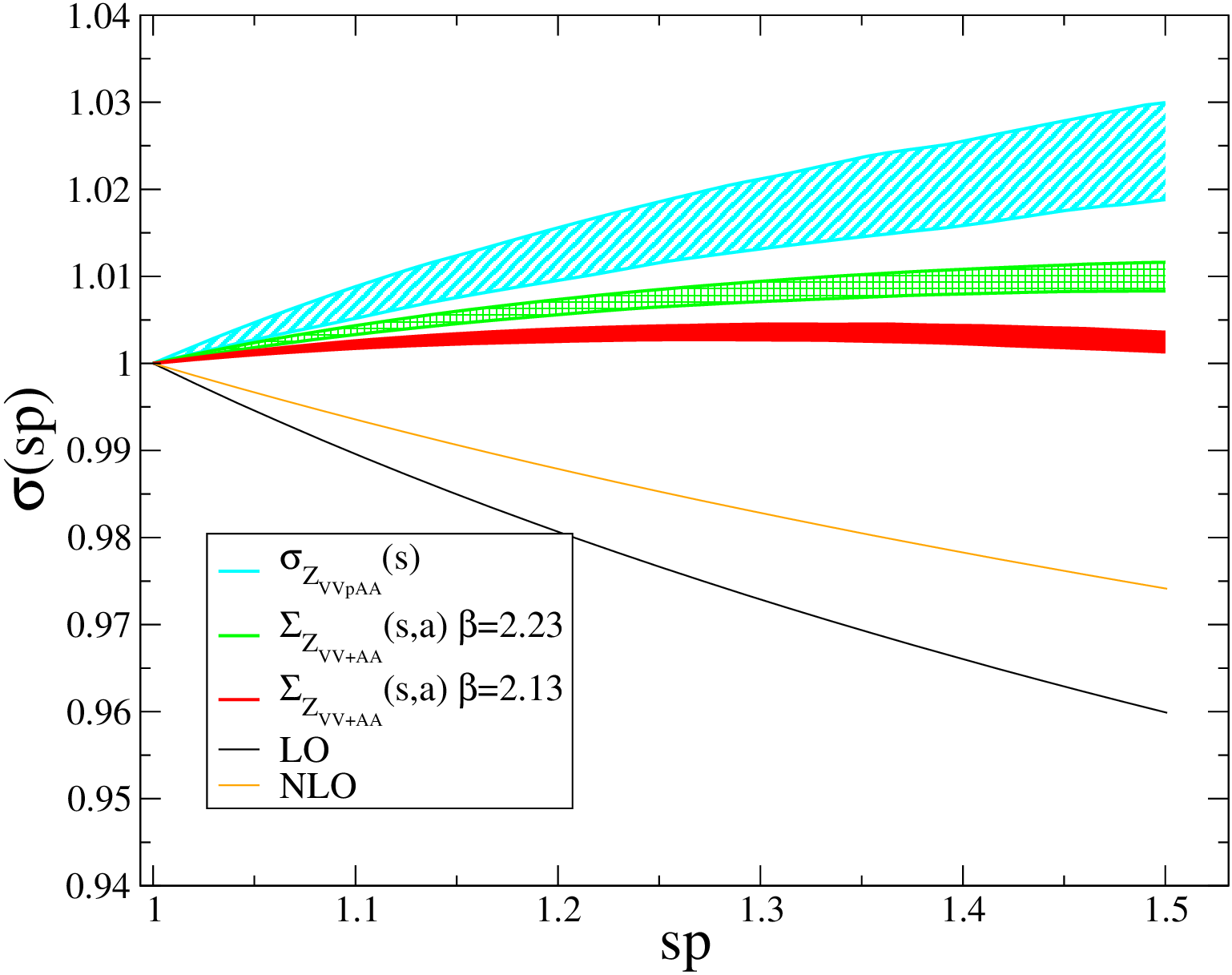}} 
\subfigure[$Z_{BK}$, non-exceptional, $SMOM-({\gamma_\mu,\cancel{p}}$)]{\includegraphics*[angle=0,width=0.45\textwidth]{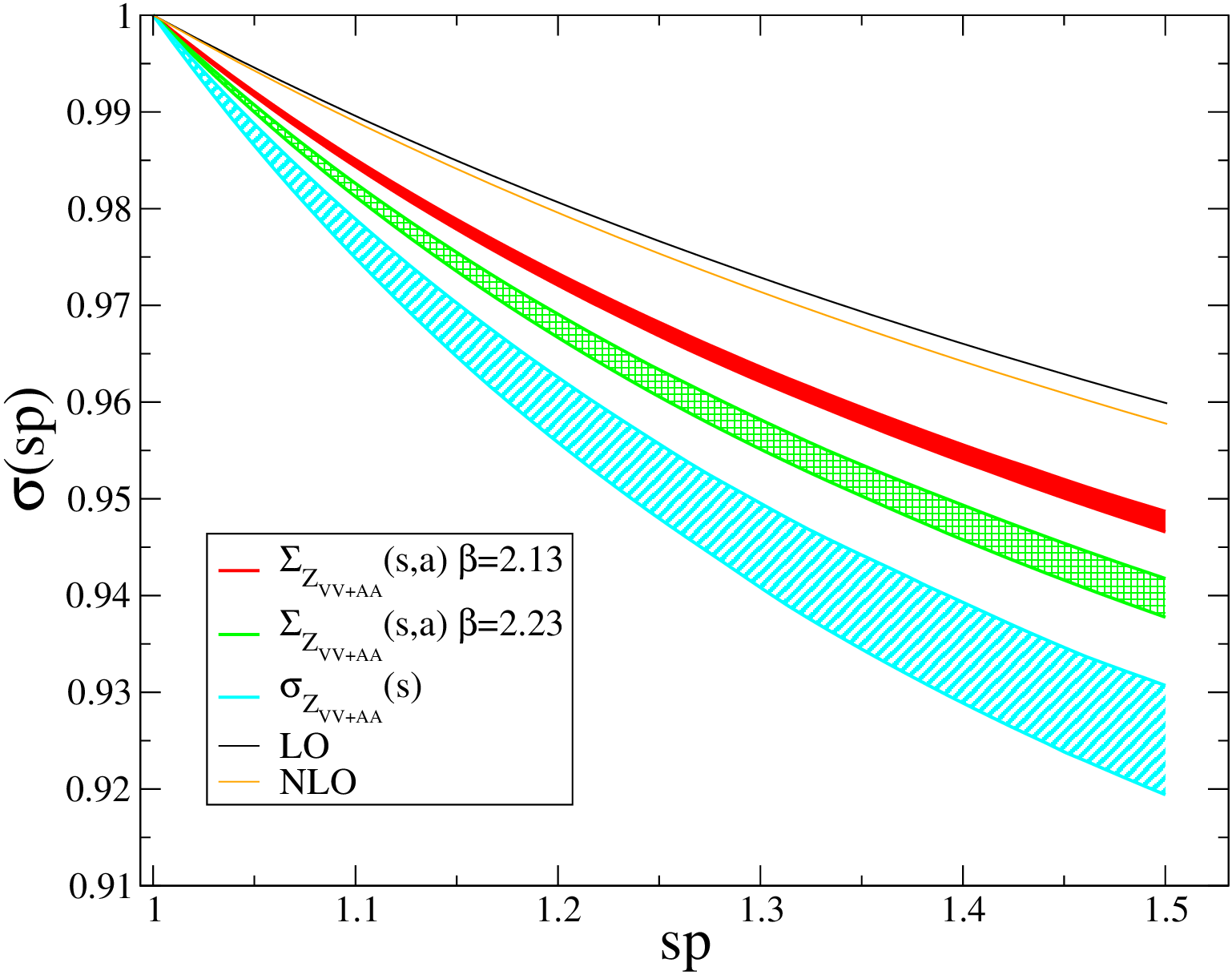}} 
  \end{center}
  \caption{Quark field renormalization running from $\simeq 2$ GeV to $\simeq 3.0$ GeV in the three different schemes compared to the perturbative running in each scheme.}
  \label{fig:ZBKrun}
\end{figure}

For each operator we can now evaluate 
$\Sigma_O(p=2{\rm GeV},sp,a^{-1}=1.729 {\rm GeV})$
and $\Sigma_O(p=2{\rm GeV},sp,a^{-1}=2.14 {\rm GeV})$. Linear extrapolation in $a^2$
with only two datapoints is not as robust as one might like, but 
certainly suffices to demonstrate the method. Using this, we can obtain the continuum
limit step scaling function $\sigma_O(p=2{\rm GeV},sp)$ for
the quark field (figure~\ref{fig:Zqrun}),
mass (figure~\ref{fig:Zmrun}), tensor current (figure~\ref{fig:ZTrun}), and
the four quark operator ${\cal O}_{VV+AA}$ (figure~\ref{fig:ZBKrun}). 
With an appropriate third and smaller lattice spacing using a correspondingly smaller
volume we could similarly determine the next step, 
evolving from 3 GeV to 4.5 GeV and so on.

Figure \ref{fig:Zqrun} is especially important, because $Z_q$ suffers
from the greatest lattice artefacts for Domain Wall Fermions, and is
carefully eliminated in the NPR analysis of other operators. 

In the $\gamma_\mu$ scheme, with both exceptional and non-exceptional 
momenta, the finite lattice spacing running of $Z_q$ is in 
the \emph{opposite} direction to the 
perturbative prediction, however the running behaviour  
recovered in the continuum limit is close to perturbative. 
The determination of $Z_q$ with domain wall fermions displays the 
momentum dependence of the exponent, $\alpha(p)$, of localisation 
in the fifth dimension \cite{Shamir:1993zy}. 
As shown here, in a number of schemes, as long as an unambiguous 
continuum limit can be defined, the Domain-Wall action will then produce
the continuum scaling behaviour of $Z_q$ and 
the other renormalization constants.

\section{Conclusions}

We have introduced the use of twisted boundary conditions
in off-shell renormalisation. This enables controlled
continuum extrapolation of the relevant amplitudes and 
rigorous disentangling of lattice artefacts from perturbative
truncation errors in the method.

We outlined a step scaling approach based on the scheme that
will allow us to raise the scales at which perturbation theory
is applied. The method has been demonstrated by taking the 
continuum limit of the first step scaling function in the process.
We can certainly raise this scale from around 2 GeV to at least around 5 GeV,
and perhaps higher. The 5GeV upper scale is limited only by the range over 
which our current matching based on the static potential is likely to be 
precise. This is, in any case, the scale above which one should
consider charm and bottom quark effects. Were an appropriately precise
method for matching lattice scales (or renormalised couplings) 
available, the method could then be applied
to evolve the three flavour theory to very high momentum 
scales, of order $M_Z$.

The lattice spacing matching
strategy will be the subject of further study, however,
raising the scale at which, typically two or three loop, 
perturbation theory is applied for Rome-Southampton renormalisation 
for much important lattice phenomenology is already a significant step.
For example, a naive estimate of an $\alpha^3$ truncation error with
O(1) coefficient is reduced from 3\% to under 1\% by raising this scale
to 5 GeV.

We plan to produce continuum limit scale evolution functions
spanning the complete range of lattice operators
covering the region 2-5 GeV.
This includes all lattice bilinear and four quark operators,
bilinear operators with covariant derivatives for structure functions
and distribution amplitudes, and three quark operators relevant for 
proton decay matrix elements.

Obtaining the scaling functions in the continuum limit will enable 
calculations with any lattice action to raise the reference scale at
which operators are quoted in $\overline{MS}$ from
2 GeV to 5 GeV with substantial reduction in systematic errors.

\section{Acknowledgements}

We would like to thank our colleagues in the RBC and UKQCD
collaborations for many fruitful discussions, and particularly
Yasumichi Aoki,  Dirk Broemmel, Norman Christ, Taku Izubuchi,
Chris Kelly, Chris Sachrajda and Christian Sturm. 
We also wish to thank Luigi Del Debbio
and David Lin for useful discussions. 
We particularly thank Chris Sachrajda and Christian Sturm for access
to unpublished projectors and perturbative expressions for the
SMOM schemes for the $VV+AA$ operator.

The software used includes:  the
CPS QCD codes \\
{\tt http://qcdoc.phys.columbia.edu/chulwoo\_index.html},
supported in part by the USDOE SciDAC program; the
BAGEL \cite{Boyle:2009bagel} assembler
kernel generator for many of the high-performance optimized kernels;
and the UKHadron codes.

The authors were supported by PPARC grants
PP/D000238/1 and PP/C503154/1. PAB acknowledges support from RCUK.

\end{document}